\documentclass[useAMS,usenatbib]{mn2e}

\usepackage{graphicx}
\usepackage{natbib}

\title[X-ray embedded stars as driving sources of turbulence]{X-ray embedded stars as driving sources of outflow-driven turbulence in OMC1-S}
\author[V.M. Rivilla, J. Mart\'in-Pintado, J. Sanz-Forcada, I. Jim\'enez-Serra and A. Rodr\'iguez-Franco]{V.M. Rivilla$^{1}$\thanks{E-mail:
rivilla@cab.inta-csic.es; ryvendel@gmail.com}, J. Mart\'in-Pintado$^{1}$, J. Sanz-Forcada$^{1}$, I. Jim\'enez-Serra$^{2,3}$
\newauthor and A. Rodr\'iguez-Franco$^{1}$
\\
\\
$^{1}$Centro de Astrobiolog\'ia (CSIC-INTA), Ctra. de Torrej\'on Ajalvir, km. 4, E-28850 Torrej\'on de Ardoz, Madrid, Spain\\
$^{2}$Harvard-Smithsonian Center for Astrophysics, 60 Garden St., 02138, Cambridge, MA, USA\\
$^{3}$European Southern Observatory, Karl-Schwarzschild-Str. 2, 85748, Garching, Germany}

\begin{document}

\date{Accepted 2013 June 24. Received 2013 June 24; in original form 2013 February 18}

\pagerange{\pageref{firstpage}--\pageref{lastpage}} \pubyear{2002}

\maketitle

\label{firstpage}

\begin{abstract}
Outflows arising from very young stars affect their surroundings and influence the star formation in the parental core. Multiple molecular outflows and Herbig-Haro (HH) objects have been observed in Orion, many of them originating from the embedded massive star-forming region known as OMC1-S. The detection of the outflow driving sources is commonly difficult, because they are still hidden behind large extinction, preventing their direct observation at optical and even near and mid-IR wavelengths. With the aim of improving the identification of the driving sources of the multiple outflows detected in OMC1-S, we used the catalog provided by deep X-ray observations, which have unveiled the very embedded population of pre-main sequence stars.  We compared the position of stars observed by the Chandra Orion Ultra Deep project (COUP) in OMC1-S with the morphology of the molecular outflows and the directions of measured proper motions of HH optical objects.
We find that 6 out of 7 molecular outflows reported in OMC1-S (detection rate of 86$\%$) have an extincted X-ray  COUP star located at the expected position of the driving source.  In several cases, X-rays detected the possible driving sources for the first time. This  clustered embedded population revealed by Chandra is very young, with an estimated average age of few 10$^{5}$ yr. It is also likely responsible for the multiple HH objects, which are the optical correspondence of flows arising from the cloud.  We show that the molecular outflows driven by the members of the OMC1-S cluster can account for the observed turbulence at core-scales and regulate the star formation efficiency. We discuss the effects of outflow feedback in the formation of massive stars, concluding that the injected turbulence in OMC1-S is compatible with a competitive accretion scenario.

\end{abstract}


\section{Introduction}

During their first stages of evolution, young stars accrete material along the equatorial plane to grow up, and spell mass in the direction perpendicular to their circumstellar disk forming bipolar outflows (\citealt{arce07}). The ubiquitous presence of outflows from low and intermediate mass stars has important implications in the formation of stars (\citealt{stanke07}), especially in the crowded clusters where massive stars are born (\citealt{lada&lada03}, \citealt{rivilla13a}). 
The outflows inject momentum and energy into the interstellar medium, favoring the fragmentation of the parental core (\citealt{knee00}; \citealt{li06}; \citealt{cunningham11}), and producing turbulence (\citealt{li06,nakamura07}), which could affect the accretion rates onto the stars (\citealt{krumholz05}). This is particularly important in the case of massive stars, that need high accretion rates to form.

These outflows can be observed over a wide range of the electromagnetic spectrum.
At optical wavelengths, the features produced when the ejected material from the stars collide with the surrounding gas and dust are known as Herbig-Haro (HH) objects. 
At longer wavelengths, the outflows are traced by the emission from different molecules (\citealt{arce07}; \citealt{beuther07}; \citealt{zapata10}; \citealt{beltran12}).

While outflows are often easily detected, however it is much more difficult to unambiguously associate them to a driving source. This is because the objects responsible for these outflows are very young stars, often hidden behind large extinction that commonly prevents their observation at optical and even near and mid-IR observations. However, X-rays can penetrate deeply into the molecular cloud, and therefore sensitive X-ray observations with high spatial resolution are a unique tool to detect most of the embedded driving sources of molecular outflows and HH objects arising from crowded stellar clusters. 
Low mass pre-main sequence (PMS) stars exhibit an enhanced magnetic activity with respect to more evolved stars, which produces bright X-ray emission. More massive stars also emits X-ray emission, usually related to wind shocks (\citealt{favata03}). 

We focus this work on identifying the driving sources of the numerous outflows detected in the massive star-forming region OMC1-S. This region, located at d$\sim$414 pc, is related the Orion Molecular Cloud (OMC) and the optically visible Orion Nebula Cluster (ONC) ionized by the four main sequence massive Trapezium stars\footnote{\citet{odell09} proposed that the OMC1-S core is detached from the background OMC, approximately at the same depth as the Trapezium stars.}.
The OMC1-S region contains a very dense and obscured molecular condensation with a mass $\sim$100 M$_{\odot}$ (\citealt{mezger90}) and a density of $n\sim$10$^{6}$ cm$^{-3}$ (\citealt{mundy86}). It exhibits several evidences of massive star formation: warm and dense gas (\citealt{ziurys81}, \citealt{batrla83}, \citealt{rodriguez-franco98}), large dust column densities (\citealt{keene82}), and presence of H$_{2}$O masers (\citealt{gaume98}).


\begin{figure*}
\vspace{1.2cm}
\hspace*{-0.5cm}
\includegraphics[angle=0,width=9.5cm]{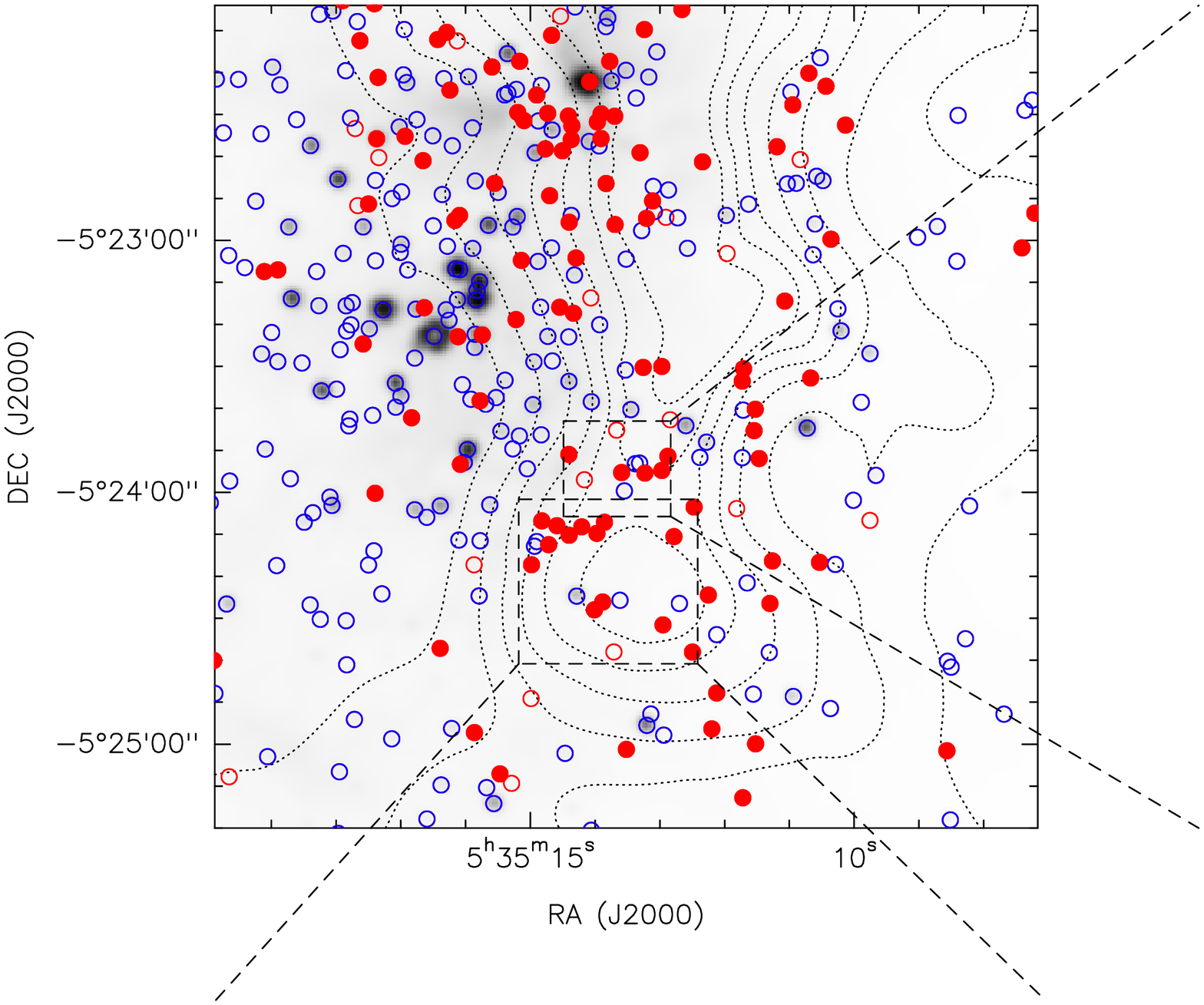}
\includegraphics[angle=0,width=8.4cm]{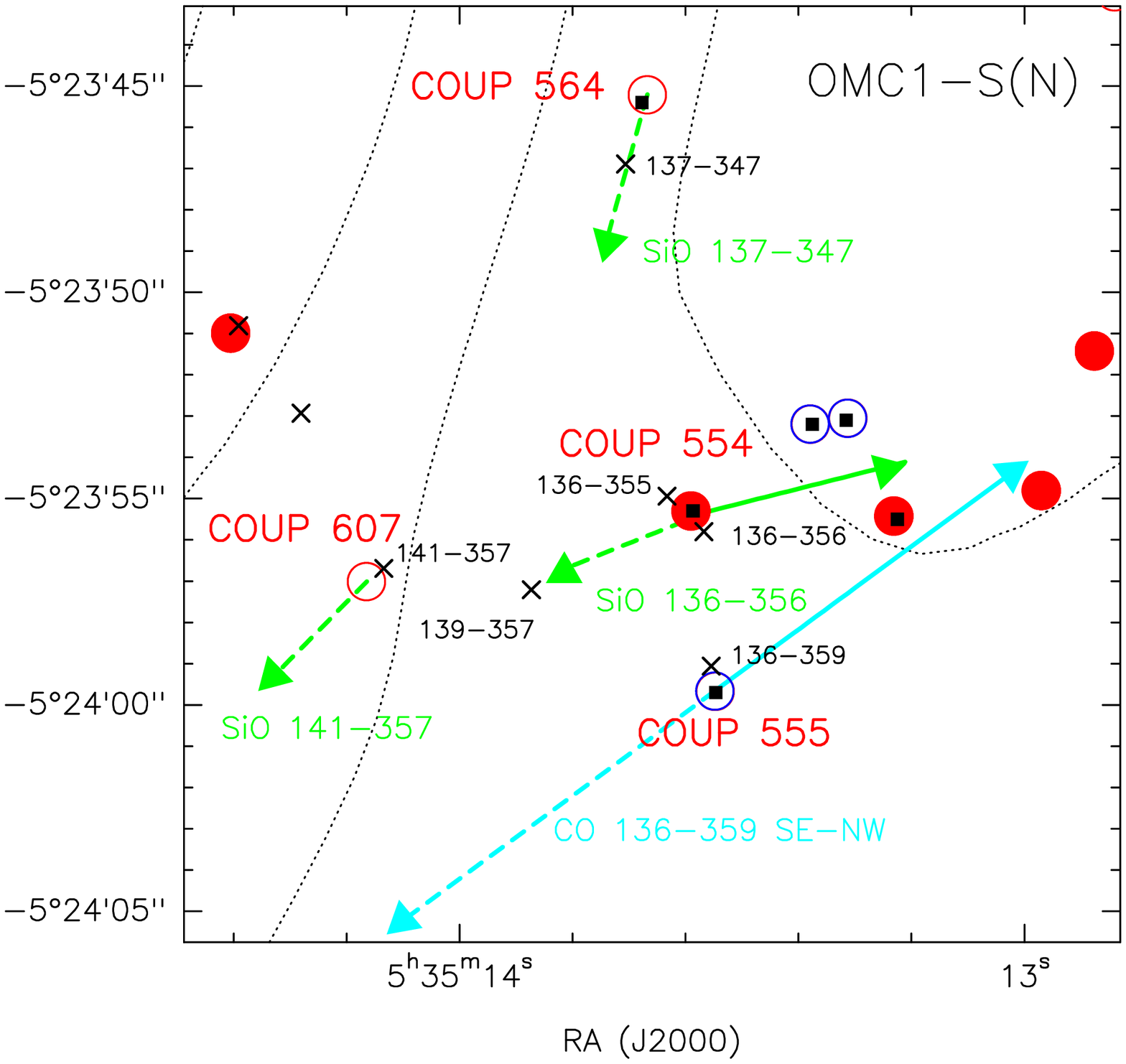}
\hspace*{-1.5cm}
\includegraphics[angle=0,width=10.0cm]{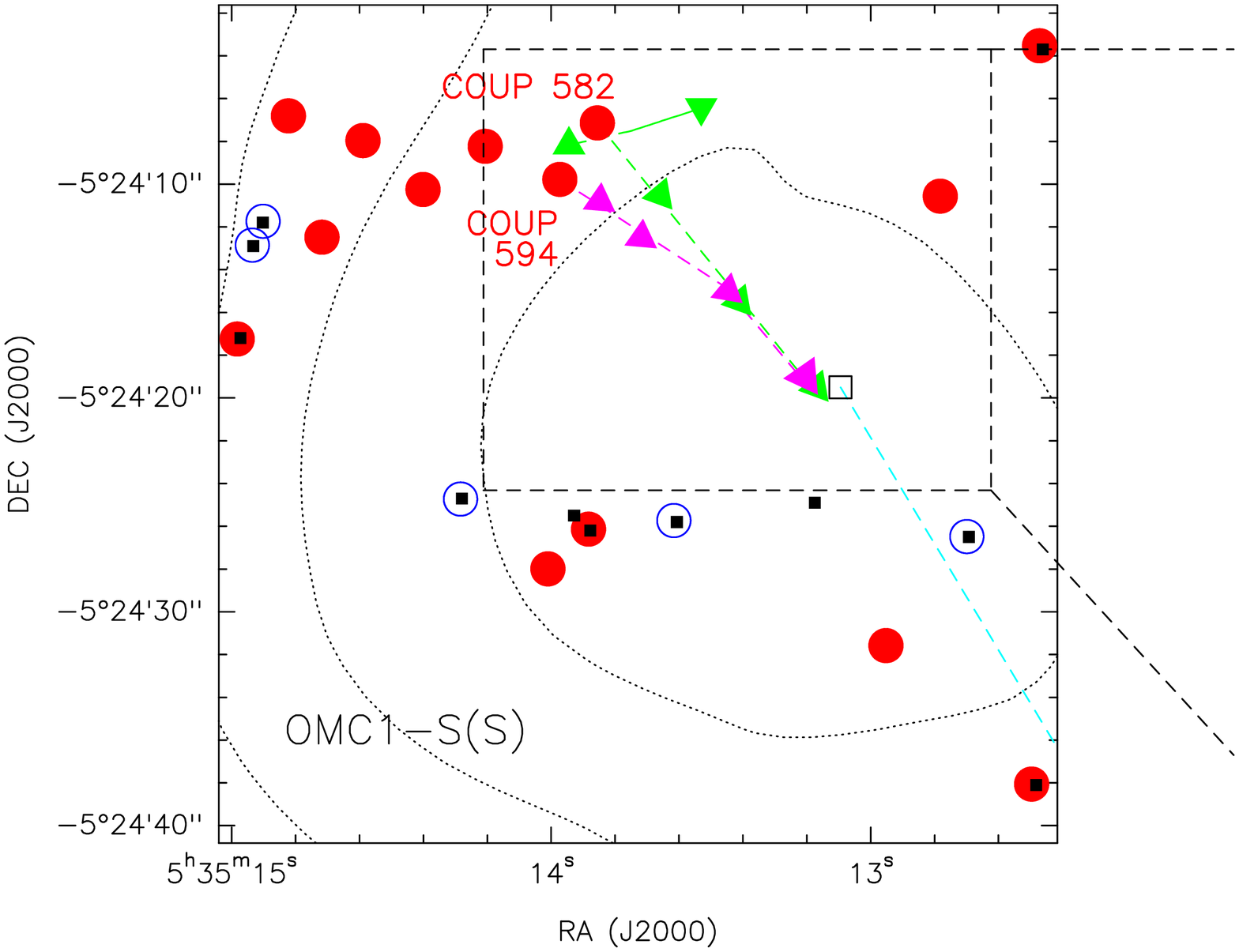}
\includegraphics[angle=0,width=7.0cm]{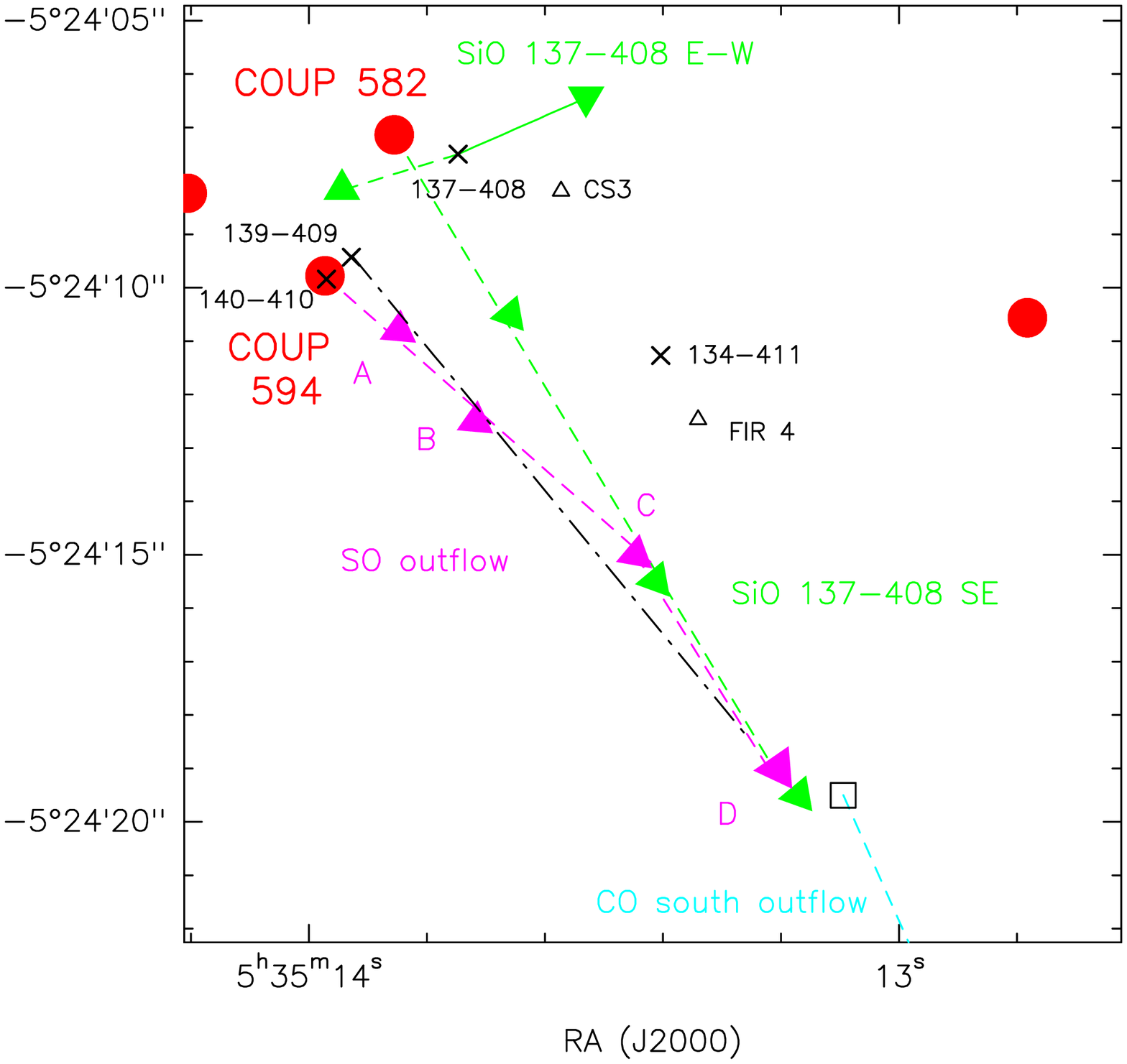}
\caption{{\em Upper Left Panel}: Spatial distribution of COUP stars in the ONC/OMC region. Open blue circles are stars with $\log N_{\rm H}<$22.5 cm$^{-2}$ (expected to be ONC members); red dots are stars with measured values of $\log N_{\rm H}>$22.5 cm$^{-2}$ (expected to be embedded members of the OMC); and open red circles are stars without measured value for $\log N_{\rm H}$ (likely also embedded objects). The gray scale is the K-band 2MASS image. The dotted contours are the emission from the OMC traced by the CN (1-0) emission, from \citet{rodriguez-franco98}. The first contour level corresponds to 7 K km s$^{-1}$ and the interval between contours is 4 K km s$^{-1}$. The dashed boxes indicate the location of the OMC1-S(N) and  OMC1-S(S) regions where several molecular outflows have been detected. 
   {\em Upper Right Panel:} Zoom-in of the OMC1-S(N) region. The CO and SiO molecular outflows from \citet{zapata05,zapata06} are indicated with arrows. The redshifted lobes are marked with dashed strokes. The COUP stars driving the molecular outflows are labeled, with the same symbols as in the upper left panel. We have also added IR stars (black filled squares; \citealt{hillenbrand00}) and cm/mm sources (black crosses; \citet{zapata04b,zapata06}.
   {\em Lower Panels:} Zoom-in of OMC1-S(S). The SiO and SO molecular outflows from \citet{zapata05,zapata06} are indicated with arrows (redshifted lobes with dashed strokes). The COUP stars have the same symbols as in the upper panels. We have also added IR stars and cm/mm sources with same notation. The triangle symbols indicate the position of FIR 4 (\citealt{mezger90}) and CS3 (\citealt{mundy86}); and the open square indicates the position from where the CO south outflow propagates in a straight line (\citealt{zapata10}).
   }
\label{figuremolecularoutflows}
\end{figure*}


The OMC1-S region shows very intense outflow activity. Three large optical flows have been detected in [OIII] ( \citealt{smith04}), along with multiple HH objects (\citealt{bally00,odell03}), and several molecular outflows (\citealt{ziurys90,schmid-burgk90,rodriguez-franco99a,zapata05,zapata06, zapata10}).

Several works (\citealt{smith04,zapata04b,robberto05,henney07}) proposed embedded mid-IR or cm/mm sources as possible candidates for driving sources of both HH objects and molecular outflows. However, none of these studies have considered the population of X-ray stars provided by the Chandra Orion Ultra-Deep Project (COUP), based on a $\sim$ 10 days observation of the ONC/OMC region. \citet{grosso05} and \citet{rivilla13a} showed that OMC1-S harbors a dense population of very embedded X-ray stars, some of them revealed by Chandra for the first time.

In this paper, we compared the position of COUP stars in OMC1-S with the morphology of the molecular outflows and the directions of measured proper motions of HH optical objects. The paper is written as follows. In Section \ref{dataanalysis} we discusse the searching procedure for the driving sources of the outflows, and study their physical properties. In Section \ref{discussion} we discuss the effects of the outflow feedback in the evolution of the stellar cluster and the formation of massive stars. In Section \ref{summary} we summarize our results and the conclusions of the paper. In Appendix \ref{appendix} we discuss extensively each outflow along with its driving source.

\section{Data analysis: Embedded COUP X-ray stars as outflow driving sources in OMC1-S}
\label{dataanalysis}

We have used the catalog of PMS X-ray stars provided by the COUP Project (\citealt{getman05a}) to compare the positions of the stars with the morphology of the molecular and optical outflows and HH objects arising from OMC1-S. In Fig. \ref{figuremolecularoutflows} we show the location of the COUP stars together with a sketch of the molecular outflows and the spatial distribution of the dense gas in the OMC.


\subsection{Membership of the COUP stellar population}
\label{sectionmembership}

One of the advantages of using X-rays to study young stellar clusters is that they suffer less foreground and background contamination than optical/IR observations. In the OMC1-S region, we expect low extragalactic contamination (\citealt{getman05b}), and low galactic contamination, because PMS stars exhibit much larger X-rays emission than older foreground stars unrelated with the cluster. Therefore, the COUP census reflect the very young stellar population of the cluster, both extremely young stars still embedded in the OMC together with more evolved (but still young) members of the optically visible ONC. To discriminate between these two populations, we follow the criteria of \citet{rivilla13a} based on the hydrogen column densities ($N_{\rm H}$) obtained from the X-ray spectra: stars with $\log N_{\rm H}<$ 22.5 cm$^{-2}$ are expected to be ONC members, and stars with $\log N_{\rm H}>$ 22.5 cm$^{-2}$ are expected to be still embedded in dense gas. 

\begin{figure}
\centering 
\hspace*{-0.5cm}
\includegraphics[angle=0,width=9cm]{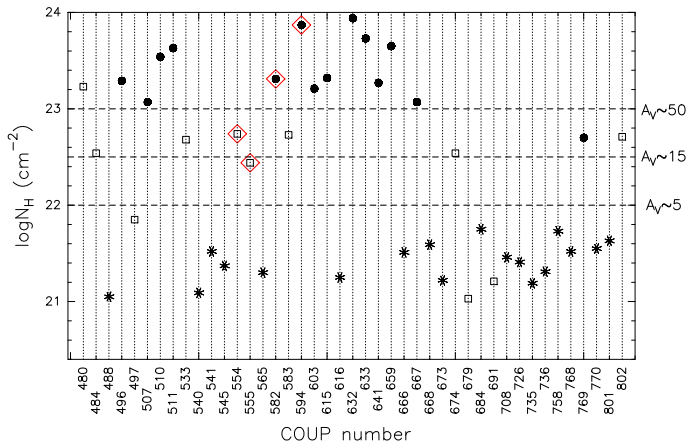}
\caption{Hydrogen column density of COUP sources located in a 60$\arcsec$ $\times$ 60$\arcsec$ region around the OMC1-S core. The plot only include those COUP stars for which $N_{\rm H}$ can be derived from the X-ray spectra. The dots are stars without optical/IR counterparts; the open squares are stars without optical counterpart but IR counterpart; and the asterisks correspond to stars with optical and IR counterparts. With red diamonds we indicate outflow-driving stars candidates.}
\label{figuremembership}
\end{figure}

Additionally, we compare here in detail the COUP census of the OMC1-S region with optical \citep{hillenbrand97} and near IR \citep{hillenbrand00} surveys. 
For our study we used a region of 60$\arcsec$ $\times$ 60$\arcsec$ around the stellar density peak found by \citet{rivilla13a}.
In Fig. \ref{figuremembership} we plot the results of our comparison between the objects detected by these surveys and the value of  $N_{\rm H}$. There is a clear trend between the presence (or not) of optical and IR counterparts as a function of $\log N_{\rm H}$.
Since the optical stars and values of log$N_{\rm H}<$ 22.5 cm$^{-2}$  are associated with ONC members, we will consider that the sources with logN$_ {H}>$22.5 cm$^{-2}$ and no optical counterpart are stars embedded in the OMC1-S core\footnote{With the exception of COUP 555 (log$N_{\rm H}$=22.44 cm$^{-2}$), that we also considered as a OMC1-S member because it drives one of the molecular outflows detected in the core (see Section \ref{sectionmolecular}) and exhibits X-ray properties typical of an embedded star (see Section \ref{sectionproperties}).}. As we will show in Section \ref{sectionproperties}, this is a good selection criteria as corroborated by our detailed analysis of the properties of the outflow driving sources in OMC1-S and of their membership to the OMCS-1 core.


Since \citet{odell09} proposed that the OMC1-S core is detached from the OMC at roughly the same depth into the ONC as the Trapezium stars, it would be possible that stars considered members of the core are actually more evolved ONC members located between the core and the background OMC. 
If we assume that these COUP stars are ONC members distributed spherically, one would expect the same number of stars in front and behind the core. However, the region zoomed-in in Fig. \ref{figuremolecularoutflows} presents $\sim$77$\%$ of extincted stars as opposed to only $\sim$23$\%$ of non-extincted stars. This deficit of non-extincted objects suggests that the extincted population is indeed located within the OMC1-S core. In fact, in a dense molecular core such as OMC1-S, with ongoing star formation and cm/mm sources and molecular outflows, one expects the presence of a stellar cluster. As found in the northern Orion Hot Core (OHC, see \citealt{rivilla13a}), the most likely scenario is that the extincted COUP stars are embedded in the core rather than being ONC members. 
Furthermore, the fact that several of the extincted COUP stars are located at the expected positions of the driving sources of the molecular outflows (Figure \ref{figuremolecularoutflows} and Section \ref{sectionmolecular}) strongly suggests that these extincted stars are indeed embedded in the core\footnote{Although we cannot completely rule out that some of the embedded objects are ONC members located between the OMC1-S core and the background OMC or a star embedded in the OMC.}.

 
\subsection{Molecular outflows}
\label{sectionmolecular}

\begin{table*}
\caption{COUP driving sources of the outflows detected in OMC1-S, with their coordinates, counterparts at other wavelengths, and other proposed candidates to drive the outflows from the literature.}            
\label{tablemolecularoutflows} 
\tabcolsep 3.pt     
\centering 
\hspace*{-0.7cm}  
\begin{tabular}{c| c| c c| c c c c c c c c | c } 
\hline 
Outflow & COUP           & RA(J2000) & DEC(J2000) & \multicolumn{8}{c}{Counterparts$^{a}$}          & Other candidates$^{d}$  \\ 
    & driving source &           &            & \multicolumn{7}{c}{IR$^{b}$}       & cm/mm$^{c}$      &                 \\
    &  &           &            & HC00 & L00 & M02 & L04 & S04 & R05 & G98         &       &                 \\
\hline
CO 136-359      & 555      & 5	35 13.55  & -5 23 59.67   & 178 & 16 & 263 & 138 & 5 & 42 & C   &     &   136-359   \\ 
SiO 136-356     & 554      & 5 35 13.59  & -5 23 55.29   & 192 & 46 & 276 & 139 & 4 & 43 & B       &     &  136-356 \\ 
SiO 137-408 SE  & 582      & 5 35 13.86  & -5 24 07.13   & -  & - & - &- & - & - & -   & -  & 137-408   \\
SiO 137-347     & 564      & 5 35 13.67  & -5 23 45.20   & 222 & - & 313 & 141 & - &-     &   -       & -  & 137-347  \\
SiO 141-357     & 607      & 5 35 14.16  & -5 23 57.01   & - & - & - &- & - & - & -   &       &  141-357    \\
SiO 137-408 E-W & -        & -           & -             & -   & - & - &- & - & -    & -  & 137-408  \\ 
SO				& 594      & 5 35 13.97  & -5 24 09.78   & -   & - & - &- & - & - &  - & 140-410		& 139-409 	\\					
CO south        & 594     &  5 35 13.97 & -5 24 09.78    & -   & -   & - & - &- & - & - &   140-410  & 139-409   \\	
HH 529        &  632    &  5 35 14.41   & -5 23 50.98     & -   &  1     &  293  &  162   &  2   &  61   & - &    144-351   & -  \\		
\hline         
\end{tabular}
\begin{list}{}{}
\item[$^{\mathrm{a}}$]{We consider as counterparts those sources with separations lower than the "counterpart radius" $r_{\rm counter}$ (see Section \ref{sectionpipo}).}
\item[$^{\mathrm{b}}$]{HC00 (\citealt{hillenbrand00}); L00 (\citealt{lada00}); M02 (\citealt{muench02}); L04 (\citealt{lada04}); S04 (\citealt{smith04}); R05 (\citealt{robberto05}); G98 (\citealt{gaume98}).}
\item[$^{\mathrm{c}}$]{1.3 cm sources from \citep{zapata04b}; and 1.3 mm sources from \citep{zapata05}.}
\item[$^{\mathrm{d}}$]{From \citet{zapata04b} and \citet{zapata05}. These cm/mm sources are located near the COUP stars, but at separations larger than the "counterpart radius" $r_{\rm counter}$ (see Section \ref{sectionpipo}).}
\end{list}
\end{table*}

In the last two decades up to seven\footnote{We do not count the CO south outflow (see lower panels in Fig. \ref{figuremolecularoutflows} and Table \ref{tablemolecularoutflows}) as a separate outflow because, as we discuss in Section \ref{collision}, it seems the result of a collision of two other outflows.} different molecular outflows have been detected in the OMC1-S region, traced by their emission in CO (\citealt{schmid-burgk90,rodriguez-franco99a,zapata05}), SiO (\citealt{ziurys90,zapata06}) and SO (\citealt{zapata10}). These outflows are in general highly collimated, which allow them to establish the expected location of the driving source. 

We compared the morphology of the molecular outflows with the stellar positions provided by the high angular resolution ($\sim$ 0.5$\arcsec$) COUP observations. In Fig. \ref{figuremolecularoutflows} we zoom-in in the two regions [ OMC1-S(N) and OMC1-S(S) ] where the molecular outflows have been detected. 
We found X-ray stars at the expected positions of the driving sources of 6 out of the 7 molecular outflows. In Table 1 we summarize the association between the outflows and the proposed COUP driving sources.

Given that the OMC1-S region exhibits a very dense population of COUP stars (\citealt{rivilla13a}), we evaluate the possibility that stars with a random distribution are located at the driving positions of the molecular outflows. The probability that members of a random distribution of $n$ stars fall within an area of size $\theta$ around the expected position of the driving source of $l$ outflows in a region with size $L$ is:

\begin{equation}
P=\frac{C_{m-l,m-n}}{C_{m,n}},
\end{equation}

where
\begin{equation}
C_{m,n}=\frac{m!}{n!\,(m-n)!}, 
\end{equation}

and $m=(L/\theta)^{2}$. Considering the population of embedded stars in OMC1-S in the two zoomed-in regions of Fig. \ref{figuremolecularoutflows}, we obtained $n=$22. The size $\theta$ is defined as a combination of the Chandra spatial resolution ($\sim$0.5$\arcsec$), uncertainties in the X-ray star positions\footnote{ \citet{getman05a} shows that the positions of COUP sources is very well determined, with values generally $<$ 0.2$\arcsec$  (see Table \ref{tabledist} for some examples).} ($<$ 0.2$\arcsec$), and the intrinsic uncertainty in the expected location of the outflow driving source. To estimate the latter, we use the maximum distance between the COUP outflow-driving candidates and other cm/mm sources candidates (\citealt{zapata04b,zapata05}). Table \ref{tabledist} shows that the maximum distance is $\sim$1.8$\arcsec$\footnote{For those outflows in which the position of the driving source is not defined with accuracy, the positional uncertainty could be somewhat larger. However, this will not change our result, i.e., that the probability of a chance alignment of all outflows with X-ray sources is very small.}. Adding quadratically the three contributions, we found a value of $\theta$=2$\arcsec$. Assuming an approximate size for the total region of $L\sim$40$\arcsec$ (0.08 pc) and 6 outflows ($l=$6) we find that the probability to find a random COUP source at the driving position of an outflow is extremely low ($P\sim$10$^{-8}$). Although we favor that X-ray stars are the driving sources (see discussion in Section \ref{sectionpipo} and Appendix \ref{appendix}), other candidates to drive some of the molecular outflows have been proposed. We repeated the calculation taking into account the worst possible case in which only 3 COUP stars are driving sources. Then, $l=$3 and the random probability is still very low ($P\sim$10$^{-4}$). As a consequence, we conclude that the X-ray stars are very likely related with the molecular outflows.

\subsubsection{X-ray sources vs. IR and cm/mm sources}
\label{sectionpipo}
 
We found COUP stars in the expected positions of the driving sources of 6 of 7 molecular outflows in the region. In this Section we compare these X-ray stars with the IR and cm/mm population detected in this region, with the aim of determining if they are counterparts of the X-ray sources, and also discuss what candidates are the most likely driving sources of the outflows.

Both cm/mm and X-rays observations are able to penetrate into the core, despite of the high extinction, which is an important advantage with respect to IR observations, which suffer much more absorption. Table \ref{tablemolecularoutflows} shows that only half (3 of 6) of the X-ray candidates have been detected at IR wavelengths. 

The comparison between the position of COUP and cm/mm sources (\citealt{zapata04b,zapata05}) provides a good opportunity to determine whether the cm/mm sources are COUP star counterparts. 
 We compared their positions with those from the IR stars (position accuracy of 0.1$\arcsec$; \citealt{hillenbrand00}). 
We consider that two sources are counterparts (i.e., their emission arises from the same object) when the separation between the sources is less than sum of the half value of their respective angular resolutions, that we call "counterpart radius" $r_{\rm counter}$. 

We present in Table \ref{tabledist} the results of this comparison for the candidates for driving sources of the outflows. We find that the COUP stars exhibit separations smaller than 0.24$\arcsec$ with respect to the IR stars. The cross-correlation of the full samples show that 95$\%$ of the sources observed in X-rays and IR have separations $<$ 0.30$\arcsec$. This is less than the "counterpart radius", $r_{\rm counter}$\footnote{Calculated from the angular resolution of COUP and IR observations from \citet{hillenbrand00}, which in both cases is $\sim$0.5$\arcsec$ in the inner region of the field of view of the observations, where the OMC1-S region is located.}, clearly confirming that the X-ray emission arise from stars. 

However, the cm sources from \citet{zapata04b} located close to IR sources (136-359, 136-356, 136-355 and 137-347, Fig. \ref{figuremolecularoutflows}) exhibit larger separations to the IR stars (0.6$\arcsec-$1.6$\arcsec$, Table \ref{tabledist}). The centimeter sources without close IR stars (141-357, 137-408 and 139-409) also present large separations (0.6$-$1.8$\arcsec$) to the closer COUP stars. Given that these separations are larger than the "counterpart radius", $r_{\rm counter}$\footnote{Using the value of angular resolution of the centimeter observations of 0.3$\arcsec$ (\citealt{zapata04b}).}=0.4$\arcsec$, it seems that these cm sources do not directly trace the position of the stars observed at X-ray and IR wavelengths.
An exception is the source 140-410, located at only 0.07$\arcsec$ from COUP 594, and that clearly is a counterpart of the X-ray star. This is confirmed by the detection of large circular polarization in this source (\citealt{zapata04b}), which is indicative of gyrosynchrotron emission from the magnetosphere of a PMS low mass star. 

In Appendix \ref{appendix} we will discuss in more detail each molecular outflow and its more likely driving source, considering the X-ray/IR and cm/mm candidates. We conclude that the COUP stars provides very good candidates for outflow drivers. X-ray emission probes efficiently the position of young PMS stars\footnote{In the COUP sample, \citet{getman05b} find that 1406 of the 1616 sources (87$\%$) are young stars associated with the stellar cluster, while only 10$\%$ are extragalactic sources and 0.1$\%$ (2 sources) are outflows emitting in X-rays.} 
, while the cm/mm continuum could also trace shocks or ionized gas from close circumstellar disks and jets (\citealt{hofner07}, \citealt{araya09}). 

Furthermore, the sensitivity of current cm/mm observations strongly limits the detection of a large fraction of the low mass stars. Recently, \citet{forbrich13} showed that in Orion the COUP observation is much more sensitive than the centimeter catalog from \citet{zapata04a}. 
Moreover, it is known that bright cm/mm emission from some stars is producing during flares (\citealt{bower03}, \citealt{forbrich08}), but it is difficult the detection outside a flare event. 

Therefore, deep X-ray observations are a very good tool, complementary to cm/mm observations, to unveil the heavily obscured young stellar population, and hence to detect the driving sources of the molecular outflows embedded in stellar clusters.

\begin{table}
\caption{X-ray position uncertainties and separations between the COUP outflow-driving stars and the nearest cm/mm and IR sources.} 
\label{tabledist}
\tabcolsep 2.pt   
\centering 
\hspace*{-0.5cm}  
\begin{tabular}{c c c c c c} 
\hline 
 COUP  & pos. unc. ($\arcsec$)$^{a}$ & $d_{\rm X-IR}$ & cm/mm &  $d_{\rm cm/mm-IR}$ $^{b}$ &  $d_{\rm cm/mm-X}$  \\ 
\hline\hline
555  & 0.04 & 0.05 & 136-359 & 0.65 & 0.61  \\
 \hline
554  & 0.01 & 0.05 & 136-356 & 0.58 & 0.61 \\
  &  &  & 136-355 & 0.78 & 0.73 \\
\hline
607  & 0.07 & - & 141-357 & - & 0.55 \\
\hline
564  & 0.11 & 0.24 & 137-347 & 1.55 & 1.78 \\ 
\hline 
582 & 0.12 & - & 137-408 & - & 1.66 \\
\hline
 594 & 0.05 & - & 140-410 & - & 0.07 \\
 &  & - & 139-409  & - & 0.76 \\
  \hline    
632 & 0.13 & 0.20 & 144-351  & 0.09 & 0.28 \\     
  
\end{tabular}
\begin{list}{}{}
\item[$^{\mathrm{a}}$]{Source's positional uncertainty of COUP sources calculated as 68$\%$ (1$\sigma$) confidence intervals using the Student's T-distribution (see \citealt{getman05a}).}
\item[$^{\mathrm{b}}$]{Sources from \citet{zapata04b} and \citet{zapata05}. The positions have been obtained from \citet{zapata04b} observations at 1.3 cm, whose angular resolution is $\sim$0.3$\arcsec$.}
\end{list}
\end{table}

\begin{figure*}
\centering 
\vspace*{-3cm}
\includegraphics[angle=90,width=18cm]{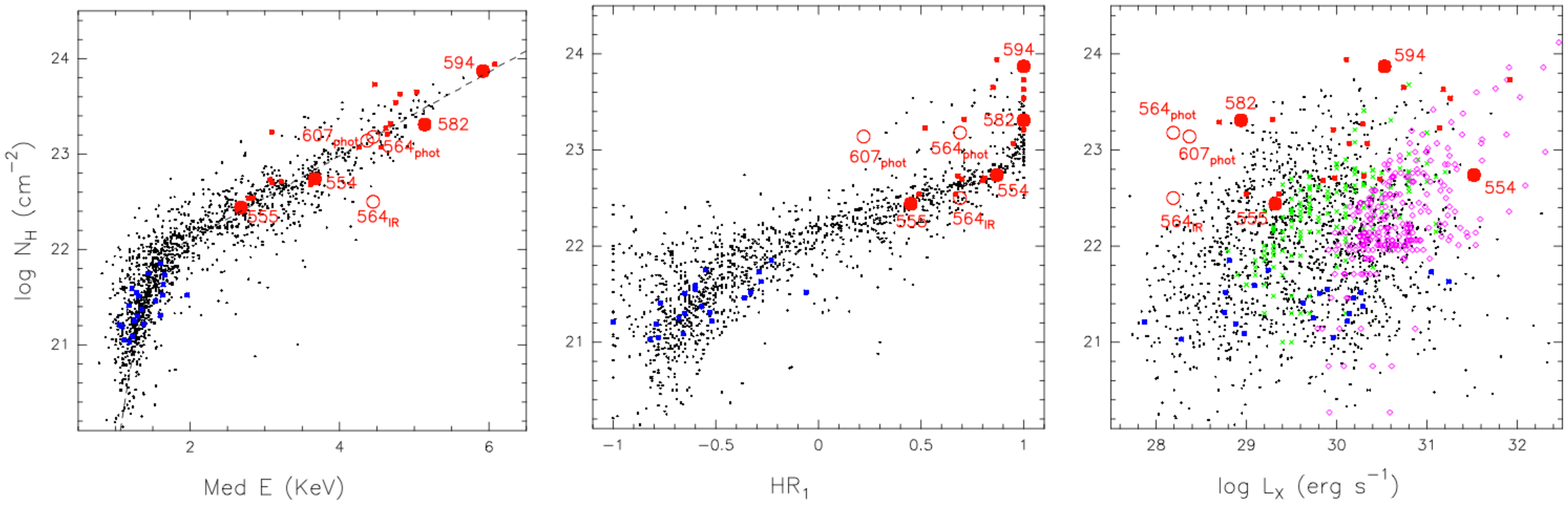}
\vspace*{-4cm}
\caption{X-ray properties of COUP stars. We plot the hydrogen column density log$N_{\rm H}$ versus: i) the mean energy of the stellar photons (left panel); ii) the hardness ratio (middle panel); and iii) the extinction-corrected X-ray luminosities (right panel). The large red circles correspond to the outflow-driving stars with values of $N_{\rm H}$ measured fitting the X-ray spectra. The large open red circles denote the COUP outflow-driving stars with $N_{\rm H}$ measured from photometry (labeled with the subscript $phot$) or from the IR counterpart (labeled with the subscript $IR$). The small red dots are the non-outflowing (but also embedded) sources of the 60$\arcsec\times$60$\arcsec$ OMC1-S region considered, while the small blue dots correspond to the non-embedded ONC members. In the {\em left} panel we also plot with a dashed line the empirical fit found by \citet{feigelson05}. In the {\em right} panel we add X-ray sources from Monoceros R2 (green crosses) and S255 (magenta diamonds), for comparison.}
\label{figureproperties}
\end{figure*}  

\subsubsection{X-ray properties of the outflow-driving stars}
\label{sectionproperties}

We discuss here more extensively the general X-ray properties of the outflow-driving stars, compared to the population of stars with no outflows (but also extincted) sources, and non-extincted ONC stars. We also considered the stellar population contained in the same region of 60$\arcsec$ $\times$ 60$\arcsec$ used in Section \ref{sectionmembership}. In Table \ref{tableproperties} we show the properties of the outflow-driving population.


\begin{table*}
\caption{X-ray properties of the outflow-driving stars in OMC1-S.} 
\label{tableproperties}
\tabcolsep 3.pt   
\centering 
\hspace*{-0.5cm}  
\begin{tabular}{c c c c c c c c c c } 
\hline 
 COUP  & net counts$^{a}$ & SNR$^{b}$ & log$N_{\rm H}$ & log$L_{\rm X}$ &  log$P_{\rm KS}$ $^{c}$ & BBNum$^{d}$ & flare$^{e}$ & $MedE$ & HR$_{1}^{f}$\\ 
\hline\hline
555  & 151 & 49 & 22.44$\pm$0.09 & 29.32 & -1.82 & 2 & (yes)  & 2.68 & 0.45$^{+0.08}_{-0.08}$   \\
554     & 14045 & 4445 & 22.74$\pm$0.01 & 31.52 & -4.0 & 31 & yes  &3.67 &  0.87$^{+0.00}_{-0.00}$   \\
582     & 18 & 9.0 & 23.31$\pm$0.48 & 28.94 & -2.32 & 1 & (yes)  & 5.14 &    1.0$^{+0.00}$    \\     
564     & 5 & 5.7 & 23.18$^{h}$ / 22.5$^{i}$ & 28.19$^{g}$ & -0.07 & 1 & - &   4.45 &  0.69$ ^{+0.38}$   \\ 
607     & 10 & 9.1 & 23.14$^{h}$ & 28.37$^{g}$ & -0.05 & 1 &- &  4.37 &    0.22$^{+0.41}_{-0.36}$    \\ 
594     & 125 & 45.0 & 23.87$\pm$0.06 & 30.53 & -4.0 & 2 & yes  & 5.92 &     1.0$^{+0.00}$   \\          
\hline
         
\end{tabular}
\begin{list}{}{}
\item[$^{\mathrm{a}}$]{Net source counts after background subtraction.}
\item[$^{\mathrm{b}}$]{The signal to noise is defined as $SNR=source$ $counts / (background$ $counts)^{1/2}$}
\item[$^{\mathrm{c}}$]{$P_{\rm KS}$ is the significance of a Kolmogorov-Smirnov (KS) test to establish whether variations are present above those expected from Poisson noise associated with a constant source (from \citealt{getman05a}). A source is considered variable if log$P_{\rm KS}<$ $-$2.0. The values have been truncated at $-$4.0.}
\item[$^{\mathrm{d}}$]{BBNum is the number of segments of the Bayesian block (BB) parametric model of source variability developed by \citet{scargle98}. A source is variable when BBNum $\geq$ 2. The values shown here are from \citet{getman05a}.}
\item[$^{\mathrm{e}}$]{From visual inspection of the X-ray light curves available at \citet{getman05a}. The parenthesis denote that the identification is uncertain.}
\item[$^{\mathrm{f}}$]{Hardness ratio $HR_{1}= (counts_{h}-counts_{s})/(counts_{h}+counts_{s})$, where subscripts $h$ and $s$ refer to the hard (2.0$-$8.0 keV) and soft (0.5$-$2.0 keV) bands, respectively.}
\item[$^{\mathrm{g}}$]{Estimated luminosity in hard band corrected for absorption, instead of luminosity in total band corrected for absorption, not available for this star.}
\item[$^{\mathrm{h}}$]{The $N_{\rm H}$ have been obtained photometrically, i.e., from the relation between the mean energy of the photons ($MedE$) and $N_{\rm H}$ found by \citet{feigelson05}.}
\item[$^{\mathrm{i}}$]{The $N_{\rm H}$ have been obtained with the relation between the color index H-K of the IR counterpart and $N_{\rm H}$, found by \citet{kohno02}.}
\end{list}
\end{table*}



In Section  \ref{sectionmembership} we established a general criterion to differentiate stars embedded in the OMC1-S core from the ONC members: no optical identification and values of $\log N_{\rm H}>$22.5 cm$^{-2}$. None of the COUP outflow-driving stars have optical counterparts. 
Regarding the values of $N_{\rm H}$, the fit of the X-ray spectra of COUP 554, COUP 582 and COUP 594 provides values of $\log N_{\rm H}>$22.5 cm$^{-2}$ (Table \ref{tableproperties}). In the case of COUP 555, its hydrogen column density is slightly lower, $\log N_{\rm H}$=22.44 cm$^{-2}$, but we also consider that this star is embedded because it clearly drives one of the molecular outflows, and its properties (see further discussion) are similar to those of the other embedded members. For COUP 564 and COUP 607 an spectral analysis was not possible given the low number of detected counts. In that case we have estimated $N_{\rm H}$ by using two alternative methods: i) from photometry, using the relation found by \citet{feigelson05} between $N_{\rm H}$ and the mean energy of the photons, $MedE$; and ii) from the correlation between the color index H-K of the IR counterpart and $N_{\rm H}$ found by \citet{kohno02}. The values obtained are log$N_{\rm H}>$22.5 cm$^{-2}$ (Table \ref{tableproperties}) and therefore all these stars are likely embedded within the OMC1-S core. 

The X-ray properties of the outflow-driving stars are seriously affected by the high extinction. \citet{feigelson05} found a relation between the median energy of background-substracted photons from the stars ($MedE$) and the hydrogen column density $N_{\rm H}$ (left panel in Fig. \ref{figureproperties}). This trend is due to an absorption effect: in the embedded sources, only the harder photons are able to escape through the molecular gas and then be detectable, while softer ones (with lower energy) are absorbed. As a consequence, the embedded stars should appear as harder sources. In the middle panel of  Fig. \ref{figureproperties} we show the values of the hardness ratio, defined as $HR_{1}= (counts_{h}-counts_{s})/(counts_{h}+counts_{s})$, where subscripts $h$ and $s$ refer to the hard (2.0$-$8.0 keV) and soft (0.5$-$2.0 keV) bands, respectively. It is clear that the embedded sources (shown in red) have much higher values than that of the foreground ONC members (blue). The high $HR_{1}$ of the sources without $N_{\rm H}$ measured from the X-ray spectra (COUP 564 and COUP 607) confirms that they are also embedded objects.

In the right panel of Fig. \ref{figureproperties} we show the values of the extinction-corrected X-ray luminosities $L_{\rm X}$. These luminosities depend on the stellar mass (\citealt{preibisch05b}), and at the age range of the stars in the region, slightly on the stellar age (\citealt{preibisch05a}). We compare the entire COUP sample (black dots) with surveys of other massive star-forming regions: Monoceros R2 (magenta diamonds, \citealt{kohno02}) and S255 (green crosses, \citealt{mucciarelli11}). In these two regions (with less sensitive Chandra observations) there is a clear trend of more luminous sources to have higher extinctions. This effect is simply due to absorption of the soft photons of sources hidden behind high extinction. In the deeper COUP sample this effect (although still present, see the absence of COUP sources in the top-left region of the right panel of Fig. \ref{figureproperties}) is less clear. 

The luminosities of the outflow-driving stars range a wide range (log$L_{\rm X}$=28$-$31 erg s$^{-1}$), similar to the values exhibited by the non-outflow-driving (but also embedded) stars and the non-embedded stars. This suggest that the differences between the luminosities are not due to the evolutionary phase of the star, but likely to the different masses of the stars.

We also study the X-ray variability. It is expected that PMS low mass stars are high variable, due to flares produced by emission from  plasma heated to high temperatures by violent reconnection events in magnetic loops in the corona. We use 3 criteria to determine the variability: i) Kolmogorov-Smirnov (KS) test to establish whether variations are present above those expected from Poisson noise associated with a constant source; ii) a Bayesian block (BB) parametric model of source variability developed by \citet{scargle98}; and iii) direct visual inspection of the X-ray light curves. The results (from \citealt{getman05a}) are shown in Table \ref{tableproperties}. The stars COUP 555, COUP 554, COUP 582 and COUP 594 fulfill at least one of this criteria, i.e., we confirm variability in a $\sim$66$\%$ of the outflow-driving stars. In the two other cases (COUP 564 and COUP 607), it is possible that the non-detection of variability is due to the low number of counts detected from these sources. Although we note that the statistic significance of this limited sample is very low, this fraction of variable stars is similar to the one shown by the non-outflowing but also embedded sample ($\sim$75$\%$) and the non-embedded sample ($\sim$80$\%$).

We conclude that there are not significant differences in the X-ray properties between the outflowing and non-outflowing (also extincted) population. The X-ray embedded sources with no outflows are also very young stars (or protostars). This reveals a population of PMS low-mass stars that outflow searches and cm/mm observations have missed so far.

\begin{figure}
\hspace{-2cm}
\includegraphics[angle=-90,width=13cm]{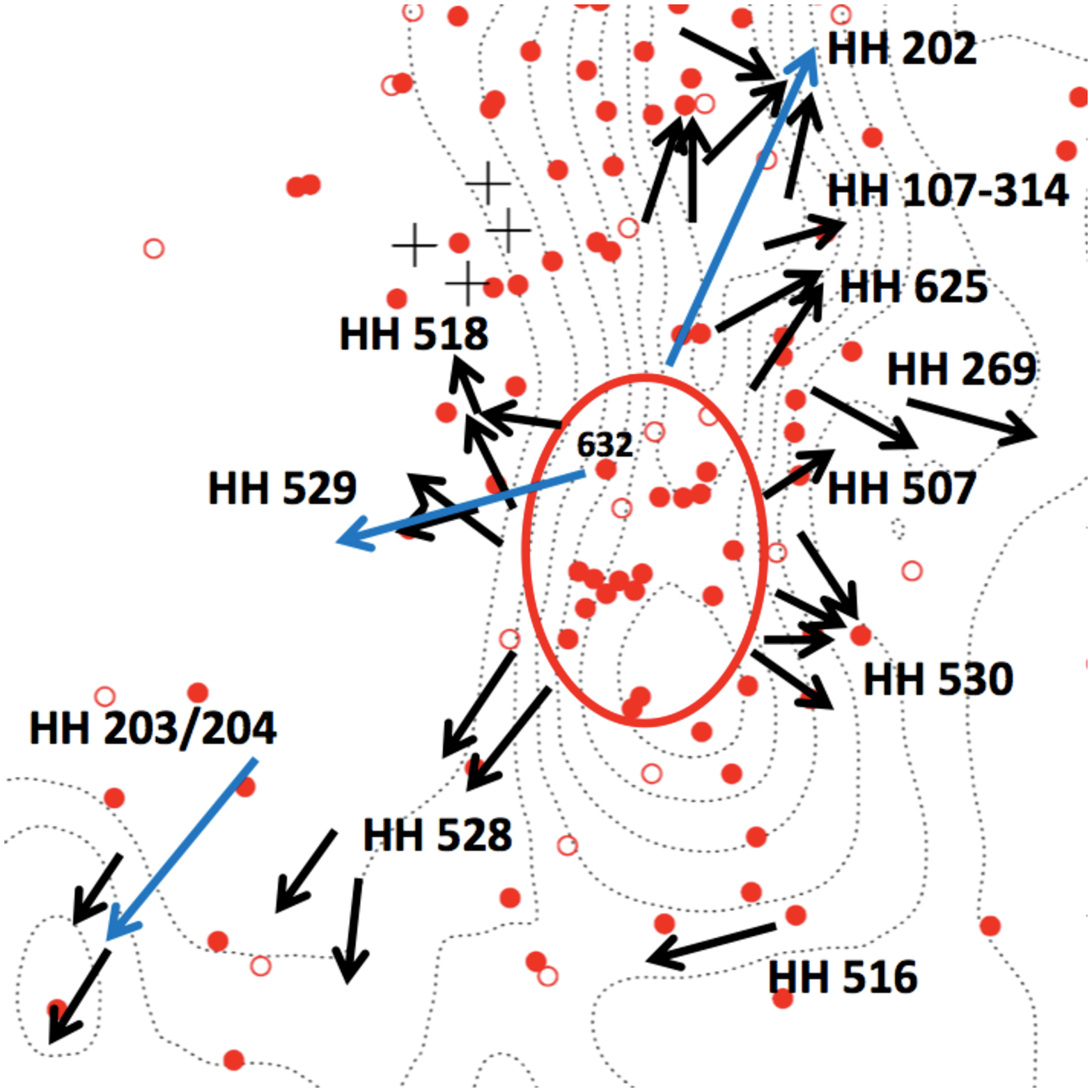}   
\caption{Distribution of the population of COUP stars embedded in the OMC. The red dots correspond to sources with $\log N_{\rm H}>$22.5 cm$^{-2}$, and the open red circles denote sources without measured value for $N_{\rm H}$ (likely also embedded objects). The orientations of the [OIII] optical outflows (\citealt{smith04}) and the main HH objects detected in the region (\citealt{bally00} and \citealt{odell03}) are indicated with blue and black arrows, respectively. We have labeled the star COUP 632 (see text). The positions of the massive Trapezium stars are indicated with 4 black crosses. The dotted contours are the emission from the OMC traced by the CN (1-0) emission, from \citet{rodriguez-franco98}. The first contour level corresponds to 7 K km s$^{-1}$ and the interval between contours is 4 K km s$^{-1}$.}
\label{hhfigure}
\end{figure}

\subsection{Optical outflows and HH objects}
\label{sectionHH}

The OMC1-S region is also the source of multiple HH objects (\citealt{bally00}, \citealt{odell03}, {\citealt{henney07}), and three large optical outflows observed in [OIII] (\citealt{smith04}). Unlike molecular outflows, whose origin can be generally well determined, in the case of HH objects it is more difficult to associate them to their driving source, because they are usually observed as the final part of a flow in the form of bow-shocks, and they are located at larger distances of their driving sources.

Although some of the HH objects might be produced by stars of the ionized nebula ONC seen at optical wavelengths, it is expected that most of them are powered by stars still embedded in the dense molecular core in OMC1-S. It is thought that the HH objects are the optical correspondence of flows which break out toward the ionized nebula ONC from the stellar population embedded in OMC1-S, where the molecular outflows are observed. Then it is worth to compare the proper motions of the HH objects (\citealt{bally00, odell03}) with the embedded population revealed for the first time by Chandra.

We show schematically in Fig. \ref{hhfigure} the distribution of embedded COUP stars, the orientations of the main HH objects (\citealt{bally00} and \citealt{odell03}), and the large optical outflows observed in [O III] (\citealt{smith04}). It is clear that the [O III] outflows and most of the HH objects point toward the same region: OMC1-S (indicated with a red ellipse in Fig. \ref{hhfigure}). 

In particular, the embedded star COUP 632 (labeled in Fig. \ref{hhfigure}) is the X-ray counterpart of the known driving source of the large optical outflow detected in [OIII] by \citet{smith04}, which is associated with multiple HH objects known as HH 529 (see Appendix \ref{appendix} for a more detailed description).

We remark the presence of the compact and dense ($>$10$^{5}$ stars pc$^{-3}$, \citealt{rivilla13a}) stellar cluster (see the region located towards the SE within the red ellipse in Fig. \ref{hhfigure}; see also the lower left panel in Fig. \ref{figuremolecularoutflows}). This cluster, which seems to drive the two molecular outflows detected in OMC1-S(S) (Section \ref{sectionmolecular}), represents a population of very young stars deeply embedded (log$N_{\rm H}>$23 cm$^{-2}$, i.e., A$_{V}>$50) that was previously undetected at other wavelengths. 

In conclusion, X-rays have detected, for the first time, numerous embedded stars that are excelent candidates to be the driving sources of the multiple HH objects found in this region.}


\section{Discussion}
\label{discussion}

In this Section we discuss the impact of the outflow feedback on the star formation in the OMC1-S core: injection of turbulence, possible disruption of the core, regulation of star formation efficiency, and role in massive star formation.

The presence of large-scale HH objects driven by OMC1-S sources (Section \ref{sectionHH}) shows that a fraction of the outflow energy and momentum is dumped outside the core. The influence of outflows in molecular clouds at parsec scales have been extensively discussed in several works (e.g. \citealt{stanke07}, \citealt{arce10} or \citealt{narayanan12}), but they did not pertain to the impact of outflows within their host cores. We focus our discussion, for the first time, on the dense material at "core-scales" ($\sim$0.05 pc). Since we are interested in the energy impact of feedback {\em inside} the core, we use the kinematics of molecular outflows, which are acting within the core, and do not include the feedback due to HH objects because the latter transfer energy {\em outside} the core.

\subsection{Timescales}
\label{timescales}

The deeply embedded stellar cluster driving molecular outflows in OMC1-S core is expected to be in a very early evolutionary stage. The molecular outflows have a typical length of $\sim$0.03 pc, which is in the low limit of sizes observed in outflows (0.03$-$2 pc, \citealt{guenthner09}). This compact sizes yield small dynamical ages (100$-$3$\times$10$^{3}$ yr, \citealt{zapata05,zapata06}). Although several uncertainties are present in these ages estimates, it is clear that these outflows are in general younger than the ones found in other star-forming regions, like L1641-N ($\sim$10$^{4}$ years, \citealt{stanke07}) or Taurus ($\sim$10$^{5}$ years, \citealt{narayanan12}). This is indicative of the youth of the OMC1-S core.

 Its compactness and the very high stellar density (\citealt{rivilla13a}) also suggest that the cluster is very young \citep{feigelson05}.
 In addition, the core shows evidence of current massive star formation, suggesting also stellar ages $\sim$10$^5$ yr. Moreover, it is known that  the earlier the evolutionary stage of the star, the larger the extinction is. \citet{prisinzano08} and \citet{ybarra13} found that stars with log$N_{\rm H}>$22.5 cm$^{-2}$ (equivalent to $A_{\rm V}>$15 mag) in Orion and Rossette Nebulae, respectively, are associated with very young objects (Class 0-I with ages of $\sim$10$^{5}$ yr; \citealt{andre94}, \citealt{evans09}). We have shown (Section \ref{sectionproperties}) that the members of the OMC1-S stellar cluster show high extinctions (log$N_{\rm H}$=22.4$-$23.9 cm$^{-2}$), suggesting that they are very young stars. In the Appendix \ref{appendix} we have fitted the spectral energy distribution (SED) of two of the outflow-driving candidates (COUP 554 and COUP 555). We obtain ages in the range [0.5$-$4] $\times$ 10$^{5}$ yr, in agreement with  \citet{prisinzano08} and \citet{ybarra13}, and supporting the youth of the cluster. Hereafter we will consider that the average age of the stellar cluster is $t_{\rm cluster}\sim$2.5$\times$10$^{5}$ yr.

Since is well established that outflows are associated particularly with the earliest stages of star formation, the presence of young stellar objects without outflows in OMC1-S may imply that star formation has been going on in the core for longer than the typical duration of the outflow phase ([0.5$-$2] $\times$10$^5$ yr, \citealt{arce10} and \citealt{narayanan12}). We can estimate this time, $t_{\rm SF}$, using the expression:


\begin{equation}
\frac{N_{\rm outflow}}{N_{\rm tot}}=\frac{t_{\rm outflow}}{t_{\rm SF}},
\end{equation}

where $N_{\rm outflow}$ is the number of stars driving outflows at present time, $N_{\rm tot}$ is the total numbers of stars in the cluster, and $t_{\rm outflow}$ is the outflow lifetime. Considering the field of view of the zoomed-in regions of Fig. \ref{figuremolecularoutflows}, the ratio  $N_{\rm outflow}/N_{\rm tot}$ is 6/22, which is $\sim$0.27. This would imply that the gas core has produced stars for a time span $\sim$4 times longer than the outflow lifetimes, i.e., of the order a few 10$^{5}$ to 10$^{6}$ yr.\footnote{This estimate is an upper limit, if we consider a little contamination by extincted ONC stars (see Section \ref{sectionmembership}) and the fact that due to confusion probably only the most prominent outflows are detected.} This 
is a strong indication that some agent has been acting to prevent the core from collapsing in its very short freefall time\footnote{$t_{\rm ff}=(3\pi)/(32G\rho)^{1/2}$, where $G$ is the gravitational constant and $\rho$ is the gas density \citep{maclow99}.}, which is $\sim$10${^4}$ Myr. In the following subsection we will show that the prime candidate for this is outflow feedback, that is able to maintain the turbulence support in the core.




\begin{table*}
\caption{Physical parameters of the OMC1-S core, and outflow feedback.}            
\label{tablefeedback}  
\tabcolsep 3.pt
\centering 
\begin{tiny}
\begin{tabular}{c c c c l l l c c c c c c c c}  
 \hline 
 \multicolumn{2}{c}{Masses (M$_{\odot}$)} & Momentum & Momentum rate &  \multicolumn{3}{l}{Vel. dispersions (km s$^{-1}$)} &  \multicolumn{2}{c}{Luminosities (L$_{\odot}$)} & \multicolumn{3}{c}{Energies ($\times$10$^{46}$ erg)} & \multicolumn{3}{c}{Timescales ($\times$10$^{5}$ yr)} \\
 
$M_{\rm core}$ & $M_{\rm esc}$ $^{a}$ & p$_{\rm outflow}$ & F$_{\rm outflow}$  & $\sigma_{\rm 1D}$ & $\sigma_{\rm gain}$ $^{b}$  & $\sigma_{\rm esc}$ $^{c}$ & $L_{\rm turb}$ $^{d}$ & $L_{\rm gain}$ $^{e}$   & $E_{\rm turb}$ $^{d}$   & $E_{\rm grav}$ $^{f}$  & $E_{\rm outflow}$  & $t_{\rm ff}$ $^{g}$ & $t_{\rm cluster}$ $^{h}$ & $t_{\rm SF}$ $^{i}$ \\

 &  & (M$_{\odot}$ Km s$^{-1}$) &  (M$_{\odot}$ Km s$^{-1}$ yr$^{-1}$)  &  &  &  &  &  &  & &   &  &  &  \\

\hline 
100 & 28 & 117 & 0.43 & 1.3 & 1.2 & 1.8 & 2.2 & 78 & 0.48 & 1.6 & 7.5 & 0.24 & 2.5 & 10 \\
\hline         
\end{tabular}
\end{tiny} 
\begin{list}{}{}
\begin{scriptsize} 
\item[$^{\mathrm{a}}$]{Mass that could potentially escape assuming that the current total outflow momentum is used to accelerate gas to reach the core escape velocity, $M_{\rm esc}=\frac{p_{\rm outflow}}{v_{\rm esc}}$, where $v_{\rm esc}=\sqrt{\frac{2GM_{\rm core}}{R_{\rm core}}}$.}
\item[$^{\mathrm{b}}$]{Gas dispersion provided by the presence of outflows, $\sigma_{\rm gain}=\frac{p_{\rm outflow}}{M_{\rm core}}$.}
\item[$^{\mathrm{c}}$]{Calculated from the escape velocity $\sigma_{\rm esc}=\frac{v_{\rm esc}}{2\sqrt{2Ln2}}$.}
\item[$^{\mathrm{d}}$]{Rate at which energy is lost due to decay of turbulence, $L_{\rm turb}=\frac{E_{\rm turb}}{t_{\rm ff}}$, where $E_{\rm turb}=\frac{3M_{\rm core}\sigma_{1D}^{2}}{2}$.}
\item[$^{\mathrm{e}}$]{Rate at which the cloud gains energy, assuming that all of the outflow momentum is converted
into turbulence in the core, and taking into account radiative losses, $L_{\rm gain}=(\sqrt{3}/2) F_{\rm outflow} \sigma_{\rm 1D}$.}
\item[$^{\mathrm{f}}$]{Gravitational energy of the core of gas, $E_{\rm grav}=\frac{GM_{\rm core}^{2}}{R_{\rm core}}$.}
\item[$^{\mathrm{g}}$]{Free-fall time, $t_{\rm ff}=\sqrt{\frac{3\pi}{32G\rho}}$.}
\item[$^{\mathrm{h}}$]{Average cluster age, calculated from the derived stellar ages of COUP 554 and COUP 555 (see Appendix \ref{appendix}).}
\item[$^{\mathrm{i}}$]{Time that the core has been forming stars.}
\end{scriptsize} 
\end{list}
\end{table*}


\subsection{Outflow-driven turbulence}
\label{sectionturbulence}

Numerical simulations by \citet{nakamura07} and \citet{carroll09} revealed that the outflows from a cluster of young stars can sustain the turbulence of the cloud gas. The detection of a dense stellar cluster and several molecular outflows in OMC1-S makes this region an excellent laboratory to discuss these effects at core size-scales ($\sim$0.05 pc). In Table \ref{tablefeedback} we present the outflows parameters (mass, momentum, momentum rate, energy, luminosity), compared with relevant magnitudes used to weight the importance of molecular outflows in the evolution of the gas core and stellar cluster.

First, we evaluate whether the energy and momentum injected by the OMC1-S outflows can potentially replenish the turbulence in the region, that decays on timescales comparable to the core free-fall time $t_{\rm ff}$.
The rate at which energy is lost due to decay of turbulence is (\citealt{stanke07}):

\begin{equation}
\centering
L_{\rm turb}=E_{\rm turb}/t_{\rm ff},
\end{equation}

with

\begin{equation}
\centering
E_{\rm turb}=(3/2)M_{\rm core}\,\sigma_{\rm 1D}^{2},
\end{equation}

where $\sigma_{\rm 1D}$ is the observed one dimensional velocity dispersion of the gas in the core and $M_{\rm core}$ the core mass.

The OMC1-S core has a current mass $M_{\rm core}\sim$100 M$_{\odot}$ (\citealt{mezger90}).
Using a radius of $\sim$25$\arcsec$ ($\sim$0.05 pc; \citealt{batrla83}; \citealt{mundy86}), we obtain that the gas density is $\rho\sim$1.3$\times$10$^{-17}$ g cm$^{-3}$ (i.e. n $\sim$ 3.5$\times$10$^{6}$ cm$^{-3}$) and the free-fall time is $t_{\rm ff}\sim$2$\times$10$^{4}$ yr.

We evaluate the 1D velocity dispersion using the linewidth of dense molecular tracers (CN from \citealt{rodriguez-franco98} and NH$_{3}$ from \citealt{batrla83}) of $\Delta v\sim$ 3 km s$^{-1}$ observed toward the region. For a 1D Maxwellian distribution of velocities, the relation between the linewidth and the one dimensional velocity dispersion is $\Delta v=2\sqrt{2\,Ln2}\,\sigma_{\rm 1D}$. Therefore, using $\sigma_{\rm 1D}\sim$1.3 km s$^{-1}$, we obtain that the rate at which energy is lost due to decay of turbulence is $L_{\rm turb}\sim$2 L$_{\odot}$.

On the other hand, the rate at which the cloud gains energy, assuming that all of the outflow momentum is converted into turbulence in the core, and taking into account radiative losses, is (\citealt{stanke07}): 

 \begin{equation}
\label{Lgain}
\centering
L_{\rm gain}=\frac{1}{2} \dot M_{\rm core}\,\sigma_{\rm 3D}^{2}=\frac{\sqrt{3}}{2}F_{\rm out}\,\sigma_{\rm 1D} ,
\end{equation}

where $\dot M_{\rm core}$ is the mass set in motion by outflows per unit time and $F_{\rm out}$ is the outflow momentum rate. Using the measured values for the molecular outflows\footnote{We have considered the 136-359 CO outflow detected by \citet{zapata05} and all the SiO outflows from \citet{zapata06}. We do not use the SO outflow and the southern CO outflow because \citet{zapata10} did not present its parameters.} ($F_{\rm out}\sim$ 0.4 M$_{\odot}$ km s$^{1}$ yr$^{-1}$) we obtain that $L_{\rm gain}\sim$ 80 L$_{\odot}$, much higher than the rate of lost energy due to decay of turbulence. This shows that the outflows are able to maintain the turbulence in the core, having a direct impact on the velocity dispersion of the gas.

The gas dispersion provided by the presence of outflows can be estimated assuming that the total momentum of the outflows ($P_{\rm outflows}\sim$ 100 M$_{\odot}$ km s$^{-1}$ for OMC1-S$^{1}$) is transferred to the core:

 \begin{equation}
\label{sigmagain}
\centering
\sigma_{\rm gain}=\frac{P_{\rm outflow}}{M_{\rm core}} ,
\end{equation}

We obtain that $\sigma_{\rm gain}\sim$ 1 km s$^{-1}$, which is equal to the observed velocity dispersion $\sigma_{\rm 1D} $. Therefore, these calculations show that the outflows driven by the members of the dense OMC1-S cluster could explain the observed turbulence in the core, and therefore they may influence any future star formation, including massive stars.

\subsection{Future core disruption}
\label{disruption}
Several works (\citealt{myers88}, \citealt{arce10}, \citealt{narayanan12}) have evaluated the outflow impact in the possible disruption of the parental cloud at parsec scales. We study here the influence of outflow feedback on the possible disruption of the more compact and dense core. One way to evaluate this possibility is by comparing the measured energy of outflows ($E_{\rm outflows}$) with the gravitational energy of the core ($E_{\rm grav}$). From Table \ref{tablefeedback}, it can be seen that in OMC-1S the ratio $E_{\rm outflows}/E_{\rm grav}$ is 4.6, which is significantly higher than the ones found by \citet{arce10} in Perseus ($<$0.40) and of the order of the one found in L1551 dark cloud in Taurus (\citealt{narayanan12}). This means that the outflows in OMC1-S could potentially disrupt the core.

Another way to evaluate the disruptive effects of outflows on the parental core is by using the velocity escape $v_{\rm esc}$ (or in terms of dispersion velocity, $\sigma_{\rm esc}=\frac{v_{\rm esc}}{2\sqrt{2Ln2}}$), which is defined as the velocity needed for the gas to escape from the gravitational potential well of the core. Table \ref{tablefeedback} shows that $\sigma_{\rm esc}$ is slightly higher than the observed dispersion velocity $\sigma_{\rm 1D}$. However, given that the values of the two velocities are close, it is possible that a fraction of the gas can be locally accelerated by outflows until it reaches $\sigma_{\rm esc}$, escaping from the core. Therefore, the molecular outflow feedback could potentially expel out material from the core. 
To estimate an upper limit of the mass that can be expelled by the molecular outflows we use the "escape mass" parameter ($M_{\rm esc}$) defined as the mass that could potentially escape the core assuming that the current total outflow momentum is used to accelerate gas to reach the core escape velocity (\citealt{arce10}):

\begin{equation}
\label{Mesc}
\centering
M_{\rm esc}=\frac{p_{\rm outflow}}{v_{\rm esc}} .
\end{equation}

In OMC1-S this escape mass is $\sim$28 M$_{\odot}$ (i.e., $\sim$28$\%$ of the current mass of the core). This suggest that if the outflow feedback is maintained throughout the entire lifetime of the core, it could potentially disrupt the gas core. We note however that a fraction of the gas will be accreted by the nascent stellar cluster before suffering ejection by outflows. \citet{matzner00} found that 50$\%$ $-$ 70$\%$ of the mass of the parental core of low-mass stellar clusters will be dispersed by outflows, while the rest (i.e., 30$\%$ $-$ 50$\%$) will be turned into stars. In Section \ref{rolestarformation} we evaluate the ability of Bondi-Hoyle accretion to transfer mass into the OMC1-S stellar cluster. 

\subsection{Outflows regulating the star formation efficiency}
\label{sectionsfe}

\citet{nakamura07} showed that the outflow turbulence leads to a delay of the gravitational collapse and, consequently, yields lower star formation efficiency (SFE), which is defined as:

\begin{equation}
\centering
SFE=\frac{M_{*}}{M_{\rm core}+M_{*}+f\times M_{outflow}} ,
\end{equation}

where $M_{*}$ is the total mass of the stellar cluster, $M_{\rm core}$ is the mass of the core, $M_{\rm outflow}$ is the total outflowing mass and $f$ is a correction factor depending on the age of the star-forming region. Assuming a value $f$=2 typical of very young regions (\citealt{arce10}), an average mass for the clusters members of 0.5 M$_{\odot}$ (as in \citealt{jorgensen08}, \citealt{evans09}, and \citealt{arce10}), we obtain a SFE$\sim$0.12, which is within the typical values found in clusters, 0.1-0.3 (e.g., \citealt{lada&lada03}). 

Given that the current star formation is dependent on the free-fall time of the core ($t_{ff}$) and the period that the region has been forming stars ($t_{\rm SF}$), we also use the "normalized" star formation efficiency defined by \citet{arce10}, $SFE_{\rm n}=SFE\times(t_{\rm ff}/t_{\rm SF})$. They found values for $SFE_{n}$ in the range 0.3$-$3.4$\%$ in a sample of 6 young star-forming regions with molecular outflows. In OMC1-S, considering the upper limit $t_{\rm SF}\sim$1$\times$10$^{6}$ yr (Section \ref{timescales}), we obtain a lower limit for $SFE_{\rm n}$ of $\sim$0.003 (i.e., 0.3$\%$). One could think that regions with high ratio between the energy rate injected by outflows and the turbulent dissipation rate ($L_{\rm gain}$/$L_{\rm turb}$) should have a very low $SFE_{n}$. However, \citet{arce10} detected no correlation between outflow strength and the $SFE_{n}$. In the case of OMC1-S, although the outflow feedback is high ($L_{\rm gain}$/$L_{\rm turb}\sim$36, Table \ref{tablefeedback}), the SFE is similar to those found by \citet{arce10}, where the impact of outflows is lower. This suggests that the outflow-driven turbulence does not prevent star formation, and it may indicate that: i) although high levels of turbulence can prevent accretion in well defined directions (the polar axis of the outflows, \citealt{cunningham06}), other paths are still available to allow accretion; ii) shocks associated with outflows compress the gas and can trigger collapse and further star formation. 


\subsection{Role of outflow feedback from a low/intermediate mass star cluster in massive star formation}
\label{rolestarformation}

Most massive stars are born within stellar clusters. Therefore, understanding cluster formation is  essential for understanding massive star formation. \citet{rivilla13a} have recently shown that massive stars in Orion are forming in very dense clusters of low mass stars; one of them is the OMC1-S cluster. In this work we have shown that members of this cluster are the driving sources of several molecular outflows. In a region of $\sim$25$\arcsec$ $\times$ 25$\arcsec$ (0.05 pc $\times$ 0.05 pc) we have detected 6 embedded X-ray stars which power coeval molecular outflows. Namely, there are at least $\sim$10$^{5}$ stars pc$^{-3}$ in this region driving outflows. 

It has been suggested that the presence of outflows influences massive star formation in several ways: i) facilitating accretion via radiation beaming along the outflow axis (\citealt{krumholz09,cunningham11}); ii) inducing fragmentation of the parental core (\citealt{knee00,li06,cunningham11}); iii) injecting turbulence in the gas environment (\citealt{li06,wang10}); and iv) triggering new stellar formation due to shock compression of the gas.

These effects have important implications for the current theories of massive star formation: monolithic core accretion (\citealt{yorke02}; \citealt{mckee02,mckee03}; \citealt{krumholz05} and \citealt{krumholz09}) and competitive accretion in stellar clusters (\citealt{bonnell06}). 

The monolithic-accretion theory needs the existence of a massive core supported against self-gravity by its internal turbulence. This turbulence can be provided by the outflows, which are able to keep the core close to virial equilibrium (\citealt{li06}). 
The massive core would collapse monolithically, producing a single massive object or a few massive stars, rather than many low mass stars (\citealt{krumholz06,krumholz09}). However, observations have shown that the molecular cores where massive stars form usually exhibit gas dispersions that prevent the virialization of the core (\citealt{martin-pintado85}). Furthermore, it is known that massive stars are found in dense clusters of low mass stars. 
Several works claimed that massive cores should fragment into many stars (\citealt{dobbs05}; \citealt{clark06}; \citealt{bonnell06}; \citealt{federrath10}; \citealt{peters10}), especially with the presence of outflows, which are believed to favor the fragmentation of the parental core, producing the formation of a cluster of low mass stars (\citealt{knee00,li06,li10,cunningham11}). 

The competitive accretion theory relies precisely in the existence of a low mass star cluster in the massive star cradles.
According to this theory, the core fragments into many condensations that form a low mass star cluster. These stars fall into local gravitational potential wells forming small-N clusters. This overall cluster potential well funnels gas to the potential center, where it is captured via Bondi-Hoyle accretion by the "privileged" stars located there, which can increase their masses.

Recently, \citet{rivilla13a} have proposed a scenario for massive star formation where the parental core fragments in multiple smaller condensations that, via monolithic accretion at "subcore" scales give birth to a low mass stellar cluster, in agreement with the observations. Once the cluster is formed, the stars that win the accretion competition can become massive via Bondi-Hoyle accretion. The efficiency of this accretion mechanism has been questioned by \citet{krumholz05}, who proposed that the turbulence of the core $-$ favored by the presence of outflows $-$ could increase the velocity dispersion of the gas and hence decrease the accretion rates.
However, \citet{bonnell06} remarked that using the local properties of massive star-forming regions, Bondi-Hoyle accretion is indeed effective to form massive stars. In agreement with this, \citet{rivilla13a} have shown that competitive accretion could be the possible mechanism for massive star formation in the Orion Hot Core.

\begin{figure}
\centering 
\includegraphics[angle=0,width=8cm]{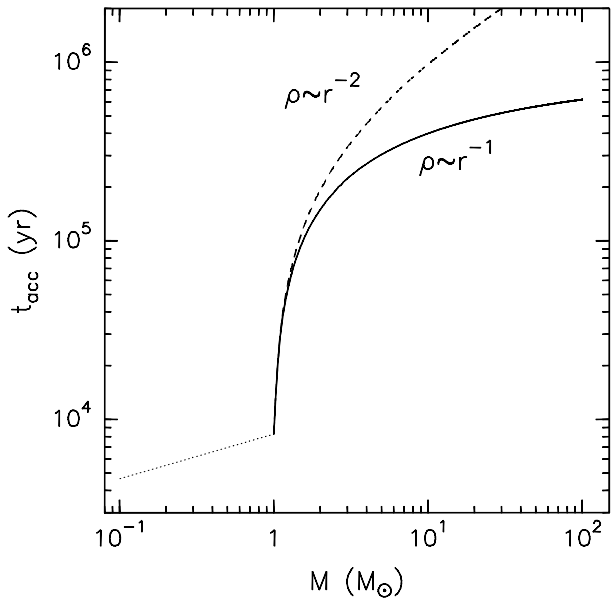}      
\caption{Accretion time to form a massive star in the OMC1-S core. The dotted line corresponds to the time needed to form the stellar seeds of 1 M$_{\odot}$ via subcore accretion, using the expression from \citet{krumholz12} for the mass accretion rate. The dashed and solid lines correspond to the accretion time needed to reach the final mass M, assuming that the gas density is $\rho\sim$r$^{-2}$ and $\rho\sim$r$^{-1}$, from the expressions presented in \citet{bonnell06} and \citet{rivilla13a}, respectively. We have used a value for the relative velocity of the gas of $v_{\rm RMS}$=2.2 km $s^{-1}$, using eq. 3 of \citet{rivilla13a} and the typical linewidth measured in OMC1-S, $\Delta\,v\sim$3 km $s^{-1}$. The volume and surface densities used are $\rho$=1.3$\times$10$^{-17}$ g cm$^{-3}$ and $\Sigma$=3.5 g cm$^{-2}$.}
\label{bondihoyle}
\end{figure}

A key parameter to determine if Bondi-Hoyle accretion is viable forming massive objects $-$ besides the gas density $\rho$ $-$ is the relative velocity of the core gas, $v_{\rm RMS}$ . As we have discussed in Section \ref{sectionturbulence}, the gas velocity can be explained by the turbulence injected in the core through the molecular outflows powered by the cluster of low/intermediate mass stars. The relative velocity of the gas can be estimated from the measured linewidth of dense gas tracers using eq. 3 from \citet{rivilla13a}, obtaining that $v_{\rm RMS}\sim$2.2 km s$^{-1}$.
Considering initial "stellar seeds" of $M = 1$ M$_{\odot}$ formed via subcore accretion, we  show in Fig. \ref{bondihoyle} the timescale for Bondi-Hoyle accretion to form a massive star. 
In order to evaluate the timescale to form the stellar seeds (dotted line in Fig. \ref{bondihoyle}) we have used the expression from \citet{krumholz12} for the mass accretion rate (with a surface stellar density of $\Sigma$=3.5 g cm$^{-2}$, obtained from the gas density $\rho$ assuming a circular geometry). The Bondi-Hoyle accretion time has been calculated from \citet{rivilla13a} and \citet{bonnell06}, who considered that the density gas has a radial dependence of $\rho(r)\sim$ r$^{-1}$ and $\rho(r)\sim$ r$^{-2}$, respectively (solid and dashed lines in Fig. \ref{bondihoyle}). This range in the radial dependence of the gas density is usually measured in massive star-forming regions (\citealt{choi00}). In our calculations we considered the mean gas density\footnote{The accretion timescale is inversely proportional to the gas density (eq. 2 from \citealt{rivilla13a}). The density decays with time (due to accretion and ejection processes), and therefore it was higher at the very early evolution and it will be lower in the future. We consider here an average value equal to the current density, that provides an average mass accretion rate.} of the core, $\rho\sim$1.3$\times$10$^{-17}$ g cm$^{-3}$.  Fig. \ref{bondihoyle} shows that the dispersion injected by outflows does not prevent that Bondi-Hoyle accretion could transfer masses $\sim$10 M$_{\odot}$ to the stellar cluster, allowing the formation of massive stars in a timescale of several 10$^{5}$ yr.

In addition, we need to take into account the inability of the outflows to inject their momentum into the whole solid angle. \citet{cunningham06} claimed that the effect of the outflows is spatially limited to the outflow axis: only those regions of the core that have been swept over by the outflow will be strongly affected. This means that the turbulence is mainly injected along the outflow axis, while the rest of the solid angle remains nearly unaffected by the energy injection, allowing the accretion in these directions. \citet{li06} remarked that it may be difficult for collimated outflows to prevent material from falling toward the center of the potential well in all directions. Therefore, the accretion of mass to the star will be oriented through the plane perpendicular to the outflow axis, i.e., the plane of the circumstellar disk.

Furthermore, it is worth to consider the impact on the turbulence of a phenomena that is rarely explored: outflow collisions. In the dense stellar clusters where massive stars form (like the one in OMC1-S) the probability of an encounter between two outflows is non negligible. In the following Section \ref{collision} we show that there is evidence of an outflow collision in OMC1-S, and we discuss its implications for the formation of massive stars.

 
\subsection{Effects of outflow collisions}
\label{collision}

Fig. \ref{figuremolecularoutflows} shows that two molecular outflows arising from the very dense stellar cluster located in OMC1-S(S) (the SO and SiO 137-408 SE outflows) have likely collided. Given that the presence of outflows is ubiquitous in star-forming regions, and that it is believed that every star might power an outflow during their formation process, it seems likely that two outflows could encounter in dense clusters like the one found in OMC1-S.
 
We can estimate the probability of a collision between two coeval molecular outflows with the expression:
\begin{equation}
\centering
P=( n_{*}\, \pi R^{2} L )^2,
\end{equation}
where $R$ and $L$ are the radius and length of the outflow, respectively, and n$_{*}$ the number density of stars. 

\citet{beltran12} already detected an encounter between two molecular outflows in the star-forming region IC 1396N. They calculated a probability of encounter of two outflows in this cluster of 5$\%$ ($P\sim$0.05).
In OMC1-S(S), the dense concentration of stars where the colliding outflows are originated has a stellar density of n$_{*}\sim$ 7.0$\times$10$^{5}$ stars pc$^{-3}$ \citep{rivilla13a}. With the mean values of the outflows ($L\sim$15$\arcsec$ and $R\sim$1.25$\arcsec$), this gives $P\sim$ 0.17, which means that the effect of collisions should be appreciable, which is consistent with the observed collision of the SiO and SO outflows. 

The lower right panel of Fig. \ref{figuremolecularoutflows} shows that the SO outflow knots do not follow a straight path, but they have a bending trajectory. We suggest that the SO outflow originates in COUP 594 and, when it reaches knot C, the outflow deviates leading to the formation of knot D. From knot C, both SO and SiO outflows share the same path, as if these two flows may have interacted. The SiO outflow could inject energy into the SO outflow, explaining the deflection in its trajectory. The initial orientation of the SO outflow could gradually change by the interaction with the SiO outflow. This is consistent with the fact that the SiO outflow is slowing down as it moves away from its driving source (\citealt{zapata06}). This energy loss would be used to deviate the SO outflow.

This interaction between both outflows might explain the detected rotation around the SO outflow axis. \citet{zapata07} found asymmetries (see their Fig. 3) in the velocities of the outer SO knots (C and D). However, they claimed that the rotation is much less evident or absent in the innermost knots A and B. This may indicate that the outflow initially does not exhibit rotation or a very slight rotation, until it finds the SiO outflow, at knot C, where the tangential forces produced by the SiO outflow induce rotation in the SO gas.

\citet{murphy08} studied the encounter of two outflows where the interference clearly affects their subsequent evolution. In their study one of the outflows is engulfed into the second one, as observed in OMC1-S. However, \citet{murphy08} considered a close binary in their simulations with a separation of 45 AU, while the separation of the driving sources of the SO and SiO outflows is $\sim$1300 AU.

\citet{cunningham06} simulated the collision of two outflows, finding that the interaction, contrary to generate more turbulence in the region, reduces indeed the redirected outflow ability to transfer momentum into the core and set gas
in motion. This is because the gas compression produced by the radiative collision shock remains within a more spatially confined region than if the outflows do not collide. As a consequence, one would expect that the collision of outflows could decrease the level of turbulence with respect to that without encounters. This would favor the Bondi-Hoyle accretion from the distributed gas component of the core, specially in the center of the cluster, where the higher stellar density facilitates the collisions and where the accretion is higher due to the gravitational attraction of the potential well. This would facilitate that the low-mass "privileged" stars could accrete enough mass to become massive. Obviously, only a fraction of the outflows in the cluster would collide, and the total impact on the core turbulence and core evolution is probably minor, although more theoretical efforts are needed to properly evaluate this effect.


\section{Summary and conclusions}
\label{summary}

In this work we correlated the OMC1-S census of stars observed in X-rays by Chandra with the expected positions of the sources driving the molecular outflows and measured proper motions of HH optical objects. 
Deep X-rays observations penetrating into the core and tracing the positions of very obscured stars are a very good tool to detect the driving sources of embedded molecular outflows. 
We find that X-ray stars are very good candidates for the driving sources of 6 out of 7 molecular outflows detected in OMC1-S (identification rate of $\sim$ 86$\%$). 
The driving sources of these outflows are very young ($\sim$10$^{5}$ yr) low/intermediate mass stars deeply embedded in the OMC1-S core, with visual extinctions of $A_{\rm V}>$ 15 mag. Moreover, X-rays observations revealed for the first time numerous embedded stars in the core that likely contribute to drive the HH objects produced when the flows break out into the ionized ONC.

Our findings stress the important role that low/intermediate mass star clusters can have on the star formation and evolution of clusters. We found that the strong molecular outflow feedback account for the observed overall turbulence in the core. This prevents its collapse in the very short free-fall time and allows to form stars during a longer period of $\leq$10$^{6}$ yr. The turbulence injected by the stellar cluster does not hinder subsequent star formation, allowing a value of the star formation efficiency in the typical range of other star-forming regions.
 
We also evaluate the impact of outflow feedback in the formation of massive stars. Outflows induce fragmentation of the parental core, favoring the formation of low and intermediate mass objects, but preventing the formation of massive stars directly from the collapse of a massive core. We show that the injection of turbulence produced by the young stellar cluster allows that Bondi-Hoyle accretion gather enough mass in a timescale of a few 10$^{5}$ yr to form massive stars.

\section*{Acknowledgments}
We thank the referee, Dr. Thomas Stanke, for his critical reading of the original version of this paper and his helpful suggestions. This work has been partially funded by MICINN grants AYA2010$-$21697$-$C05$-$01 and  FIS2012$-$39162$-$C06$-$01, and Astro$-$Madrid (CAM S2009/ESP$-$1496), and CSIC grant JAE$-$Predoc2008. I.J$-$S. acknowledges the funding received from the People Programme (Marie Curie Actions) of the European Union's Seventh Framework Programme (FP7/2007$-$2013) under REA grant agreement number PIIF$-$GA$-$2011$-$301538.

\bibliographystyle{mn2e}
\bibliography{bib-omc1s}

\appendix

\section{Individual comments on outflows and driving sources}
\label{appendix}

\begin{table*}
\caption{Physical parameters obtained from the fit of the SEDs of COUP 555, COUP 554 and COUP 632 (driving sources of the molecular outflows CO 136-359 and SiO 136-355, and the HH 529 optical flow, respectively).}            
\label{tableappendix}  
\tabcolsep 3.pt
\centering 
\begin{tabular}{c c c c c| c c c c c c c}  
 \hline 
COUP    & $A_{\rm V}^{1}$  &&& & \multicolumn{7}{c}{SED results} \\

star  & (mag)  &&& & $M$ (M$_{\odot}$) & $L_{\rm bol}$ (L$_{\odot}$) & Stellar age ($\times$10$^{5}$ yr) & $A_{\rm V}$ (mag) & log[$L_{\rm X}$/$L_{\rm bol}$] & $M_{\rm disk}$ (M$_{\odot}$) &  $M_{\rm envelope}$ (M$_{\odot}$)\\
 \hline 
555 & 14 &&&  & 0.3$-$2 & 3$-$10 & 0.5$-$4 & 15$-$25 & [-4.7,-5.3] & 0.01$-$0.05 & 0.02$-$2 \\
554 & 28 &&& & 0.9$-$2.5 & 8$-$30 & 1.5$-$4 & 25$-$30 & [-3.0,-3.5]  & 0.03$-$0.1 & 0.004$-$3 \\
632 & $>$400 &&& & 0.5$-$1.5 & 7$-$50 & nwc$^{2}$ & 40$-$60 & [-4.3,-5.2] & 0.0002$-$0.002 & 0.01$-$0.5 \\
\hline         
\end{tabular}
\begin{list}{}{}
\item[$^{\mathrm{1}}$]{From the value of log$N_{\rm H}$ obtained fitting the X-ray spectra, and using the \citet{vuong03} relation.}
\item[$^{\mathrm{1}}$]{Parameter not well constrained.}
\end{list}
\end{table*}

In this Appendix we discuss the details of the different molecular outflows located in OMC1-S, along with their proposed driving sources. We also include the large optical flow associated with HH 529, whose driving source has also been observed in X-rays (COUP 632).

\begin{figure*}
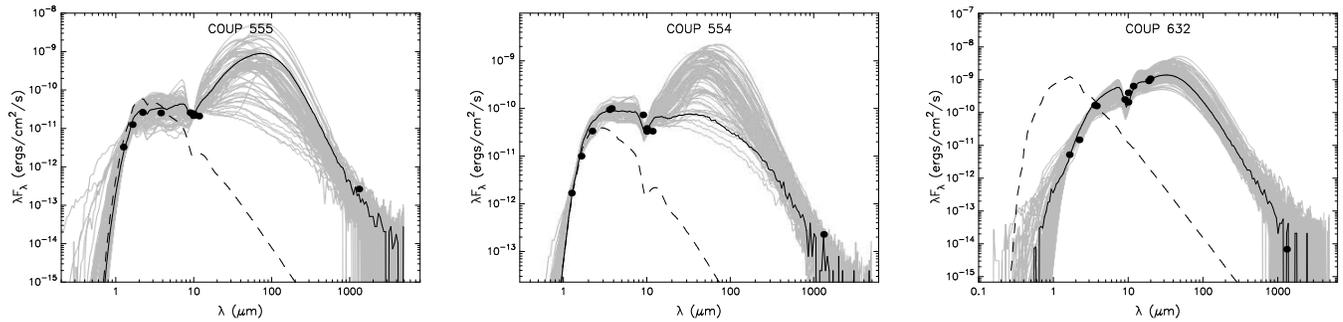

\centering
\includegraphics[angle=0,width=5.5cm]{COUP555-SED-100.eps}
\hspace{0.4cm}  
\includegraphics[angle=0,width=5.5cm]{COUP554-SED-100.eps}  
\hspace{0.4cm}  
\includegraphics[angle=0,width=5.5cm]{SED-632.eps}  
\caption{SEDs of COUP 555 (left), COUP 554 (middle) and COUP 632 (right) obtained from Robitaille et al. (2006, 2007) models. The filled circles show the input fluxes. The black lines show the best fit, the gray lines show the 100 best fits, and the dashed lines show the stellar photosphere corresponding to the central source of the best fitting model, as it would look in the absence of circumstellar dust (but including interstellar extinction).}
\label{figureSED}
\end{figure*}

For the stars COUP 554, COUP 555 and COUP 632 (with detections at IR wavelengths) we fitted its spectral energy distribution (SED) with the models from \citet{robitaille06,robitaille07}.
These stars are presumably very young, and thus they likely have a gas envelope responsible for the SED in the mid-IR to millimeter range. Unfortunately, there are not available mid/far-IR measurements at wavelengths $>$20 $\mu$m. In the millimeter range, to constrain the envelope contribution we used the fluxes of the 1.3 mm sources 136-356, 136-359 and 144-351 detected by \citet{zapata05} (see Table \ref{tabledist}). The results of the 100 best fits are shown in Fig. \ref{figureSED}. Although the SED is not well characterized in the mid/far-IR range due to the lack of observational data, we can derive a range for the physical parameters (stellar mass, total luminosity, stellar age, extinction, fractional luminosity [$L_{\rm X}/L_{\rm bol}$], disk mass and envelope mass; Table \ref{tableappendix}). In all cases, we selected the parameter range shared by more than 50$\%$ of the best models (see Fig. \ref{figureSEDluminosity}, where we show for illustrative purposes the results for the bolometric luminosity of COUP 554 obtained considering the best 100 models).

In the case of COUP stars in which we cannot fit the SED, we used indicators as the X-ray variability, flaring activity and fractional luminosity ($L_{\rm X}$/$L_{\rm bol}$) to distinguish qualitatively between low mass (M$<$1.3 M$_{\odot}$), intermediate mass (1.3 M$_{\odot}<M<$8 M$_{\odot}$), and high mass (M$>$8 M$_{\odot}$) stars. While high X-ray variability and X-ray flares attributed to plasma trapped in closed magnetic structures and violently heated by magnetic reconnection events are common in lower mass objects, they are not expected in more massive stars. We determined the variability using the same criteria as in Section \ref{sectionproperties}.

The fractional luminosity $L_{\rm X}$/$L_{\rm bol}$ indicates the efficiency of the star in converting luminous energy into magnetic activity. 
It is expected that PMS stars with lower masses can reach the "saturation limit" for coronally active stars (emitting up to log~[$L_{\rm X}/L_{\rm bol}$]=$-$3.0; \citealt{pallavicini81,fleming95,flaccomio03,gudel09}). In the case of low/intermediate (M$<$ 3 M$_{\odot}$) mass accreting stars, \citet{preibisch05b} found that this value is slightly lower, log~[$L_{\rm X}$/$L_{\rm bol}$]=$-$3.7. This value is significantly different in stars with higher masses, more inefficient converting bolometric luminosity in magnetic energy, where log~[$L_{\rm X}$/$L_{\rm bol}$] reaches values down to $-$7.0 (\citealt{pallavicini81}). In Table \ref{tableappendix2} we show the masses of COUP stars without SED, obtained applying the canonical relations log~[$L_{\rm X}$/$L_{\rm bol}$]=$-$3.7 and log~[$L_{\rm X}$/$L_{\rm bol}$]=$-$7.0. In the next subsections we will discuss that these sources without SED are likely low/intermediate mass objects.

\subsection{The 136-359 SE-NW CO outflow (COUP 555)}

This outflow was first detected by \citet{rodriguez-franco99a,rodriguez-franco99b}, and subsequently resolved with higher angular resolution by \citet{zapata05} (Fig. \ref{figuremolecularoutflows}). The position of the driving source coincides with the X-ray star COUP 555, located at 0.6$\arcsec$ of the cm/mm source 136-359  (\citealt{zapata05}).
Based on the large mechanical luminosity of the outflow, \citet{zapata05} suggested that the driving source of this outflow could be an embedded high luminosity ($L_{\rm bol}\sim$10$^{4}$ L$_{\odot}$) and massive object. 
They used in their calculation a correlation between the mechanical luminosity of the outflow and the luminosity of the driving star (\citealt{beuther02}). However, this correlation was established from single-dish data (which considerably underestimates the luminosity due to beam dilution), and therefore it can not be used with the mechanical luminosity obtained from the interferometric data from \citet{zapata05}. Using the mass of the outflow calculated from the single-dish data from \citet{rodriguez-franco99b}, $m_{\rm outflow}\sim$ 0.04 M$_{\odot}$, we have used the single-dish correlation presented by \citet{wu04} between the outflow mass and the bolometric luminosity of the driving source, obtaining that the luminosity of the  driving star must be $\sim$1 L$_{\odot}$, which implies a low mass star.  We note that this is a rough estimate with high uncertainty, because the outflow mass considers only the very high velocity gas and the \citet{wu04} relation exhibits a large dispersion.

Other arguments can be made to show that a low/intermediate luminosity source could drive the outflow: i) there are indications that the outflow is very young (low dynamical age, very high collimation, very high CO velocities), and it is well known that
younger outflows have higher luminosity for a given driving source luminosity (e.g., \citealt{bontemps96}); and ii) the outflow is traced out to exceptionally high velocities, which will artificially boost the outflow luminosity, and mimic an outflow driven by a much higher luminosity source than it actually is.

With the aim of better constraining the mass of this outflow-driving star, we fitted their SED using the fluxes provided by several IR observations (see Table \ref{tablemolecularoutflows}; \citealt{gaume98}; \citealt{hillenbrand00}; \citealt{lada00}; \citealt{muench02}; \citealt{lada04}; \citealt{smith04}; \citealt{robberto05}) with the \citet{robitaille06,robitaille07} models. The 100 best fits are shown in the left panel of Fig. \ref{figureSED}, and the physical parameters derived are presented in Table \ref{tableappendix}. From the models, we obtain a range for the extinction of $A_{\rm V}$=15$-$25 mag, consistent with that obtained from X-rays ($A_{\rm V}\sim$14 mag). 
We obtain a stellar mass of $M$=0.3$-$2 M$_{\odot}$, a bolometric luminosity of $L_{\rm bol}=3-$10 L$_{\odot}$, and a young stellar age of (0.5$-$4)$\times$10$^{5}$ yr, as expected. The star exhibits a moderate X-ray variability (Table \ref{tableproperties}), and it may show a flare, although the large error bars in the light curve prevent from a clear confirmation. The log~[$L_{\rm X}$/$L_{\rm bol}$] ratio ranges between -4.7 and -5.3, which are more consistent with an intermediate-mass star than a low-mass star. Thus we favor the case of a mass in the higher part of the range mass resulted from the SED fit, i.e., $\sim$2 M$_{\odot}$.

\subsection{The 136-356 bipolar SiO outflow (COUP 554)}

This is the main SiO outflow detected by \citet{zapata06}, produced by the embedded star COUP 554, which has an IR counterpart (\citealt{gaume98,hillenbrand00, muench02}). \citet{hashimoto07} detected infrared polarization signatures typical of inner cavities of stellar outflows toward this position, confirming that this object is the driving source.

\begin{figure}
\centering
\includegraphics[angle=0,width=8cm]{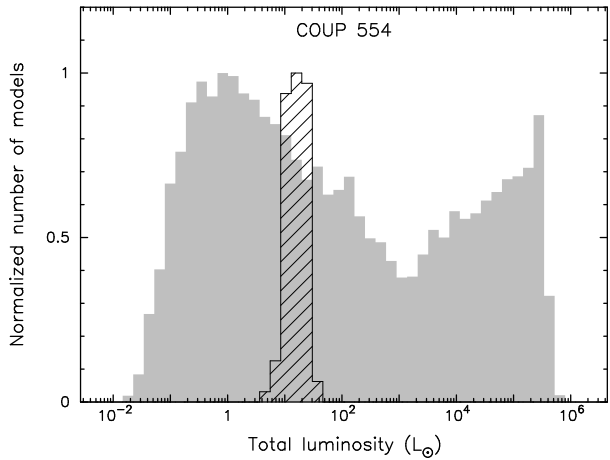}
 \caption{Results for the bolometric luminosity of COUP 554. The gray histogram shows the distribution of models in the model grid, and the hashed histogram shows the distribution of the 100 best selected models.}
\label{figureSEDluminosity}
\end{figure}

\citet{gaume98} called this object source B, and estimated the mass of the star from its IR K-band flux, finding that if it were a zero age main sequence (ZAMS) star, it would be a B2 star (with a mass $\sim$ 10 M$_{\odot}$ and a luminosity of 3$\times$10$^{3}$ L$_{\odot}$). Using the fluxes measured from different works (see Table \ref{tablemolecularoutflows}) we fitted the SED of COUP 554 with the \citet{robitaille06} models to better contrain the properties of this outflow-driving star. The right panel of Fig. \ref{figureSED} and Table \ref{tableappendix} show the results. We obtained a value for the extinction of $A_{\rm V}$=25$-$30 mag, consistent with that obtained from X-rays, $A_{\rm V}\sim$28 mag. The models give a mass of $M$=0.9$-$2.5 M$_{\odot}$, a luminosity of $L_{\rm bol}=$8$-$30 L$_{\odot}$ (Fig. \ref{figureSEDluminosity}), and a young stellar age of (1.5$-$4)$\times$10$^{5}$ yr. 
The presence of X-rays flares in its light curve, and the high X-ray variability (Table \ref{tableproperties}) also allow us to rule out the massive nature of this star. The ratio between the X-ray luminosity and the bolometric luminosity log~[$L_{\rm X}$/$L_{\rm bol}$] range between -3.0 and -3.5, consistently with a star with masses lower than 3 M$_{\odot}$, in agreement with the mass calculated from the SED.
  
The cm/mm continuum sources 136-356 and 136-355 are located at separations of 0.61$\arcsec$ and 0.73$\arcsec$, respectively, outside the "counterpart radius" $r_{\rm counter}$ (Section \ref{sectionpipo}), suggesting that the emission does not directly arise from the star\footnote{The possibility that COUP 554 is the combined counterpart to the cm sources is ruled out because the angular resolution of Chandra in the OMC1-S region ($\sim$0.5$\arcsec$) would resolve these sources, which are separated by 1.3$\arcsec$.}. The position of the cm/mm sources, approximately perpendicular to the SiO outflow axis, may indicate that they are tracing ionized gas or shocks produced in a circumstellar disk.
The cm source 139-357 (Fig. \ref{figuremolecularoutflows}) is located in the outflow redshifted lobe axis of the SiO outflow, suggesting that it is tracing shocks related with the flow.

\begin{table}
\caption{Fractional luminosities of the COUP driving sources of OMC1-S outflows without SED fitting.}            
\label{tableappendix2}  
\tabcolsep 3.pt
\centering 
\begin{tabular}{c c c}  
 \hline 
COUP star & \multicolumn{2}{c}{$M$ (M$_{\odot}$)} \\
  &    log[$L_{\rm X}$/$L_{\rm bol}$]=$-$3.7 & log[$L_{\rm X}$/$L_{\rm bol}$]=$-$7.0\\
\hline

564  & $<$0.1 & 3.5 \\  
607    & $<$0.1 & 4 \\
582   & 0.1 & 5-6 \\
594   & 0.6 & $>$7\\ 
\hline         
\end{tabular}
\end{table}


\subsection{137-347 Monopolar Outflow (COUP 564)}
This is the northernmost outflow detected by \citet{zapata06} (upper right panel in Fig. \ref{figuremolecularoutflows}). These authors claimed that it arises from the cm/mm compact source 137-347. 
This source is unresolved at 1.3 cm by the beam of the observations (0.3$\arcsec$), so we can set an upper limit for their size of 0.15$\arcsec$. Given that the separation with the X-ray star is significantively larger ($\sim$1.8$\arcsec$) than this size and than the $counterpart$ $radius$ (see Section \ref{sectionmolecular}), we conclude that these two sources are not counterparts. \citet{zapata05} detected 137-347 also at 1.3 cm, but in this case with an elongated shape of size $\sim$2.5$\arcsec$ along the same direction (SE-NW) of the outflow lobe (\citealt{zapata06}). We interpret that the star COUP 564 (also detected in IR) is the driving source of the outflow, whose innemost part is traced by the shocks that produce the continuum emission of 137-347, and outermost part is traced by the SiO lobe.


The X-ray variability of COUP 564 is not confirmed (Table \ref{tableproperties}), although this might be due to the few counts detected from this source. Using the canonical relations for stars with lower and higher masses, we obtain masses of $<$0.1 M$_{\odot}$ and 3.5 M$_{\odot}$, respectively (assuming the estimated average age of the cluster of 2.5$\times$10$^{5}$ yr and the \citet{siess00} PMS stellar models).

\subsection{141-357 Outflow (COUP 607)}

This outflow is located to the east of the 136-356 main bipolar SiO outflow (see upper right panel of Fig. \ref{figuremolecularoutflows}). \citet{zapata06} noted that this outflow does not belong to the 136-356 bipolar outflow, because: i) it is located $\sim$10$\arcsec$ away from the driving source; ii) it shows much lower velocity gas; and iii) if powered by the same object as the bipolar outflow, its dynamical age would be much older.
 \citet{zapata06} proposed that this outflow is likely powered by their radio compact source 141-357. The X-ray observations reveal the presence of a star, COUP 607, so far undetected in near or mid-IR, located at $\sim$0.55$\arcsec$ (Fig. \ref{figuremolecularoutflows} and Table \ref{tabledist}). Therefore, we support the outflow nature of this molecular component, powered by this star. As we discussed in Section \ref{sectionmolecular} the separation between the X-ray and cm sources is slightly larger than the "counterpart radius" $r_{\rm counter}\sim$0.5$\arcsec$. Given its relative location with respect to COUP 607 and the outflow axis (Fig \ref{figuremolecularoutflows}), 141-357 may trace a close jet of the blueshifted counterlobe of the observed SiO flow.

The presence of a flare in the X-ray light curve is uncertain due to poor signal-to-noise ratio, and no clear variability is detected. Using the relation for higher mass stars, we obtain a mass 4 M$_{\odot}$; although a low mass star with $\leq$0.1 M$_{\odot}$ cannot be ruled out.

\subsection{The 137-408 SE SiO outflow (COUP 582)}
\label{COUP582}
This redshifted SiO outflow was found by \citet{zapata06} in OMC1-S(S) (lower right panel in Fig. \ref{figuremolecularoutflows}). It has a clumpy structure along the outflow axis, with "knots" at different velocities. Unlike most of molecular outflows (such as e.g. L1448-mm; \citealt{guilloteau92}), this outflow is slowing down as it moves away from its origin, with moderate velocities (29$-$56 km s$^{-1}$) with respect to the ambient cloud velocity for the inner part, and low velocities (0$-$27 km s$^{-1}$) for the outer one.

\citet{zapata06} claimed that the outflow appears to be driven by the millimeter source 137-408 (labeled in the lower right panel of Fig. \ref{figuremolecularoutflows}). If this is the case, this would indicate the presence of a close binary, because 137-408 is very likely the driving source of the 137-408 E-W outflow (Section \ref{137EW}). 
However, given that the well defined outflow axis (green dashed line) passes south ($\sim$1.5$\arcsec$) of the millimeter position, we think that the very embedded X-ray star COUP 582 is a more natural candidate, because it is perfectly aligned with the outflow axis, and consequently there is no need to invoke a bending trajectory.
This star is one of the members of the densest stellar cluster detected in OMC1-S(S) (\citealt{rivilla13a}), with $\log N_{\rm H}$=23.3 cm$^{-2}$ (equivalent to $A_{\rm V}\sim$ 100 mag). It does not have cm/mm, optical or IR counterparts, being detected for the first time in X-rays.  It seems to exhibit a flare (although the large error bars in the light curve prevent a clear confirmation) and a moderate X-ray variability, suggesting that it is a low mass star. Using the relation log~[$L_{\rm X}$/$L_{\rm bol}]\sim$ $-$3.7, we obtain a mass of $\sim$ 0.1 M$_{\odot}$. However, a star with higher mass cannot be completely ruled out. Using the relation log~[$L_{\rm X}$/$L_{\rm bol}]\sim-$7, the mass would be $\sim$5$-$6 M$_{\odot}$.

\subsection{The SO outflow (COUP 594)}
\label{SOoutflow}

This redshifted outflow traced by SO (\citealt{zapata10}) is located very close to the SiO 137-408 SE outflow (see lower panels in Fig. \ref{figuremolecularoutflows}). This SO outflow exhibits four differentiated knots (A, B, C and D). 
Regarding the origin of the SO outflow, \citet{zapata10} suggested that the cm/mm source 139-409, a circumbinary ring (\citealt{zapata07}), could be the driving source. In the lower right panel of Fig. \ref{figuremolecularoutflows} we have plotted with a black dashed line the hypothetical outflow axis, with P.A.=48$\textordmasculine$. It seems that the SO knots are not well aligned. Actually, it is difficult to explain the SO outflow from the 139-409 position with a straight flow. \citet{zapata10} already recognized that a bending trajectory can not be excluded for this outflow. However, in order to produce the four knots from the 139-409 origin, it would be needed at least two bends, which although posssible, seems unlikely. Furthermore, \citet{zapata10} noted that the identification of the driving source still remained open because the orientation and sense of rotation of the 139-409 circumbinary ring does not coincide with those detected in the outflow. Furthermore, another circumbinary ring detected in the region, 134-411, with similar characteristics to 139-409 (\citealt{zapata07}) and expected to be in the same evolutionary stage, do not drive a molecular outflow (see Fig. \ref{figuremolecularoutflows}).


We consider that the X-ray star COUP 594 is a more plausible driving source. It is located at $\sim$0.7$\arcsec$ from 139-409, and coincident with the 1.3 cm source 139-409 (\citealt{zapata04b}). As discussed in Section \ref{sectionpipo} this cm source is very likely produced by gyrosynchrotron emission from the magnetosphere of the star revealed by X-rays. 

As shown in Fig. \ref{figuremolecularoutflows}, the SO outflow could be originated from COUP 594, with a P.A.$\sim$57$\textordmasculine$ (magenta dashed line) until the knot C, where the outflow deviates to P.A.$\sim$40$\textordmasculine$. The change in orientation occurs at the knot C, which is coincident with a SiO knot of the 137-408 SE SiO outflow. It is striking that both SO and SiO outflows share the same path from this point, and suggests that these two outflows may be interacting, as discussed in Section \ref{collision}.

Based on its high X-ray variability and the presence of a flare in its light curve, we conclude that COUP 594, is very likely a low mass star. Using the relation log~[$L_{\rm X}$/$L_{\rm bol}$]=$-$3.7, we obtain a mass of $\sim$ 0.6 M$_{\odot}$ (Table \ref{tableappendix2}).

\subsection{The CO south outflow (COUP 594 + COUP 582)}
\label{COoutflow}

\citet{schmid-burgk89} detected a large-scale CO outflow toward OMC1-S(S), that was  
re-observed by \citet{zapata10} with better spatial resolution. 
The powering source of this outflow is still not clear. 
From \citet{zapata10} it is difficult to determine exactly its origin, although they established a position from where it propagates in a straight line (white square in the lower right panel of Fig. \ref{figuremolecularoutflows}). The X-ray observations did not detect any source along the outflow axis.

\citet{zapata10} speculated that this CO outflow is originated at a position toward the NE, where the redshifted SO outflow is detected. They found evidence for rotation in both outflows, which may indicate that they are related. They claimed that the CO outflow could be the extension of the SO outflow. Both outflows have a similar velocity range between 8 and 25 km s$^{-1}$, and  additional observations showed that SO also extends to the SW (see Fig. 4 from \citealt{zapata10}), coincident with the orientation of the CO outflow. In addition, the SO clump D is nearly coincident with a CO clump, which suggests a link between both flows. As we have discussed in Section \ref{collision} and Section \ref{SOoutflow}, the SO outflow seems to collide with the SiO 137-408 SE outflow, which provides energy to the SO flow and produces its deviation. Therefore, and given that \citet{zapata10} did not find CO emission at the inner part of the SO outflow (before the encounter), we interpret that the CO outflow is the continuation of the encounter of the SO and SiO outflows powered by COUP 594 and COUP 582, respectively.

\subsection{137-408 E-W Outflow}
\label{137EW}
This SiO emission exhibits two lobes, one blueshifted (between $-$30 and $-$80 km s$^{-1}$) and the other redshifted (between 30 and 80 km s$^{-1}$). \citet{zapata06} suggested that the strong millimeter source 137-408 could be the driving source. \citet{zapata06} already noted that the presence of two outflows around this position (this one and SiO 137-408 SE) may indicate the presence of more than one embedded driving source. The closest X-ray star is COUP 582, shifted $\sim$1.7$\arcsec$ with respect to the origin of the 137-408 E-W  lobes. As shown in Section \ref{COUP582}, this X-ray star might drive the SiO SE 137-408 outflow. Given that the millimeter source 137-408 is located more symmetrically with respect to the lobes of the 137-408 E-W outflow, this source is the most likely driving source.

\subsection{The optical outflow HH 529 (COUP 632)}

This large optical outflow detected in [OIII] (\citealt{smith04}) is associated with the multiple HH objects known as HH 529. Its driving source is detected in X-rays (COUP 632), near-IR (\citealt{lada00,muench02,lada04}), mid-IR (source 2 from \citealt{smith04}), and also at 1.3 cm (source 144-351 of \citealt{zapata04b}) and 1.3 mm (\citealt{zapata05}). As indicated by \citet{zapata04b}, the radio emission migth emanates from the magnetosphere (gyrosynchrotron emission) or the ionized outflow (free-free emission) of a low/intermediate mass star. 

\citet{hashimoto07} found polarization signatures around this star, characteristic of cavities along the inner parts of outflows, and \citet{robberto05} detected an IR elongated filament immediately to the west of the star with the same orientation as the HH 529, supporting that it is the driving source of the optical outflow and the HH objects. 

This star is one of the three reddest stars of the entire L-band (3.5 $\mu$m) survey from \citet{lada00}, and its K-L excess suggests a visual extinction of A$_{\rm V}>$100. The high extinction is supported by the X-ray observations, because COUP 632 has the highest hydrogen column density measured in the entire COUP sample, logN$_{H}$=23.94 cm$^{-3}$. This extremely high value may also indicate the presence of an edge-on circumstellar disk. The extinction estimated from X-rays (which corresponds to A$_{\rm V}>$400 mag) is even larger than the obtained from IR. This could indicate that there is a "gas excess" around this star, may be because it is still forming and accreting material. The X-ray source does not exhibit X-ray variability, although this can be due to the low number of detected net counts (16). We fitted the SED of this star using the measured fluxes from the literature (right panel in Fig. \ref{figureSEDluminosity}). The results confirm the high extinction suffered by this object and indicates a stellar mass in the range 0.5$-$1.5 M$_{\odot}$. We favor the higher mass limit, because the obtained fractional luminosity L$_{\rm X}$/L$_{\rm bol}$ range is more consistent with an intermediate-mass star than a low-mass star (Table \ref{tableappendix}).



\label{lastpage}

\end{document}